\documentclass[aps,prb,twocolumn,superscriptaddress,showpacs,amsmath,amssymb,footinbib,longbibliography]{revtex4-2}
\usepackage[english]{babel}
\usepackage{graphicx}
\usepackage{dcolumn}
\usepackage{bm}
\usepackage{subfigure}
\usepackage[utf8x]{inputenc}
\usepackage{color}
\usepackage{textcomp}
\usepackage[colorlinks,bookmarks=false,citecolor=blue,linkcolor=red,urlcolor=blue]{hyperref}

\newcommand{\op}[1]{%
    \fontdimen12\textfont3=2pt\fontdimen12\scriptfont3=1.4pt%
    \!\null\mathop{\vphantom{#1}\smash{#1}}\limits_{\sim}\null\!}
\newcommand{\xref}[1]{\protect\ref{#1}}
\newcommand{\figref}[1]{Fig.~\protect\ref{#1}}
\newcommand{\tabref}[1]{Tab.~\protect\ref{#1}}
\newcommand{\fmref}[1]{(\protect\ref{#1})}
\def\bra#1{\langle \, {#1} \, | \,}
\def\ket#1{\, | \, {#1} \, \rangle}
\newcommand{\braket}[2]{\langle \, {#1} \, | \, {#2} \, \rangle}

\newcommand{\Tr}{\mbox{Tr}}

\renewcommand{\eqref}[1]{Eq.~(\protect\ref{#1})}
\newcommand{\secref}[1]{Sec.~\protect\ref{#1}}

\newcommand*{\Ndeg}{{N_{\text{\scriptsize deg}}}}

\begin{document}
\title{Accuracy of the typicality approach using Chebyshev polynomials}

\author{Henrik Schl\"uter}
\author{Florian Gayk}
\affiliation{Fakult\"at f\"ur Physik, Universit\"at Bielefeld, Postfach 100131, D-33501 Bielefeld, Germany}
\author{Heinz-J\"urgen Schmidt}
\affiliation{Universit\"at Osnabr\"uck, Fachbereich Physik,
 	D-49069 Osnabr\"uck, Germany}
\author{Andreas Honecker}
\affiliation{Laboratoire de Physique Th\'eorique et Mod\'elisation, CNRS UMR 8089,
CY Cergy Paris Universit\'e, F-95302 Cergy-Pontoise Cedex, France}
\author{J\"urgen Schnack}
\email{jschnack@uni-bielefeld.de}
\affiliation{Fakult\"at f\"ur Physik, Universit\"at Bielefeld, Postfach 100131, D-33501 Bielefeld, Germany}

\date{\today}

\begin{abstract}
Trace estimators allow to approximate thermodynamic equilibrium 
observables with astonishing accuracy. A prominent representative 
is the finite-temperature Lanczos method (FTLM) which relies on
a Krylov space expansion of the exponential describing the Boltzmann
weights. Here we report investigations of an alternative approach
which employs Chebyshev polynomials. This method turns out
to be also very accurate in general, but shows systematic inaccuracies 
at low temperatures that can be traced back to an improper behavior 
of the approximated density of states with and without smoothing kernel.
Applications to archetypical quantum
spin systems are discussed as examples.
\end{abstract}


\maketitle

\section{Introduction}

The (numerically) exact evaluation of thermodynamic quantum equilibrium observables is restricted to small
systems due to the exponential growth of the Hilbert space for systems with finite-size
single-site Hilbert spaces such as Heisenberg or Hubbard models. For quantum systems
with unrestricted single-site spaces the situation is even more severe. Only very few
analytically solvable systems are known which creates a massive need for numerical 
(approximation) schemes. One rather successful means to approximate thermodynamic quantities
rests on trace estimators which approximate a trace by an expectation value with respect
to a random vector 
\cite{Ski:88,Hut:CSSC89,DrS:PRL93,JaP:PRB94,SiR:IJMPC94,GoM:Stanford97,HaD:PRE00,WWA:RMP06,AvT:ACM11,RoA:FCM15,SAI:NM17,WCW:PRR19}. 
These schemes, sometimes also called \emph{typicality} or \emph{(microcanonical) thermal pure 
quantum states} \cite{IMN:IEEE19,SuS:PRL12,SuS:PRL13,OAD:PRE18},
have been used very successfully in particular in the field of correlated electron systems, see
e.g. \cite{ABK:PRB75,DRdV:ZP89,dVDR:PRB93,Dag:RMP94,ADE:PRB03,ZST:PRB06,ScW:EPJB10,USL:JMMM13,HaS:EPJB14,PHK:PRB14,SGB:PRB15,SHP:PRB16,YSY:PRB16,ScT:PR17,OAD:PRE18,SSR:PRB18,PrK:PRB18,RKK:PRB19,SSH:PRL20,LaO:NPJQM20,MoT:PRR20,HRS:CMP20}
but also in quantum chemistry \cite{MHL:CPL01,HLM:JCP02}. 

A prominent formulation of the method is the finite temperature
Lanczos method (FTLM) \cite{JaP:PRB94,JaP:AP00,ADE:PRB03,PrB:SSSSS13,PRE:COR17} which employs
a Krylov space expansion for $\exp\{-\beta\op{H}\}$ (operators are marked by a tilde).
It turns out that FTLM produces very accurate approximations 
when estimates are averaged over random vectors (order of $\sim 100$, fewer for larger spaces), 
compare 
\cite{LSY:A13,RoA:FCM15,OAD:PRE18,SSR:PRB18,SRS:PRR20,SRH:ZNA20,MoT:PRR20}.

Despite this success, the authors of \cite{WWA:RMP06} suggest that an alternative approximation
using an expansion of the density of states in terms of Chebyshev polynomials should
be more accurate \cite{WWA:RMP06}. 
The major argument is that this expansion does not suffer from the loss of orthogonality 
during recursive state generation used in Krylov space methods.
This property is certainly responsible for the high accuracy obtained in numerical unitary
time evolution using a Chebyshev expansion, 
see e.g.~\cite{IiE:PRL03,WWA:RMP06,HWM:PRB11,TMP:PRB14,TMP:PRB14B,LaZ:PRR19}.

In the present paper, we therefore study several Heisenberg 
quantum spin systems and derive numerical as well as formal conclusions about the accuracy
of the method. We can summarize that the approach via Chebyshev polynomials is indeed accurate,
but not more accurate than FTLM \cite{SRS:PRR20}. On the contrary, under certain circumstances the employed
kernel which smoothens (unphysical) oscillations of the approximated density of states introduces
systematic inaccuracies. The same holds for the mapping of the energy spectrum onto
the interval $[-1+\varepsilon/2,1-\varepsilon/2]$ to comply with the domain of definition of the polynomials.

The paper is organized as follows. In Section \ref{sec-2} we recapitulate the
Chebyshev method. In Section~\ref{sec-3} we present our numerical examples. 
The article closes with a discussion in Section~\ref{sec-4}.

\section{Method}
\label{sec-2}
In this section, we briefly introduce the Chebyshev method and 
its parameters to be able to discuss the method's accuracy. 
For a more detailed description of the algorithm we recommend \cite{WWA:RMP06}.   

In a quantum mechanical system with a discrete energy spectrum, the microcanonical 
density of states is defined as
\begin{align}\label{E-2-1}
\rho(E) := \sum_{n}\delta(E-E_n)\ .
\end{align}
The canonical partition function $Z(\beta)$ is determined by the integral over 
the density of states weighted with the Boltzmann factor:
\begin{align}\label{E-2-2}
Z(\beta)=\sum_{n}e^{-\beta E_n}=\int_{-\infty}^\infty\rho(E)e^{-\beta E}dE
\end{align}
with $\beta =\tfrac{1}{k_B T}$.
Correspondingly, the heat capacity is evaluated as 
\begin{eqnarray}\label{E-2-2-B}
\frac{C(\beta)}{k_B}=
\beta^2
\Bigg[
&&
\frac{1}{Z(\beta)}
\int_{-\infty}^\infty\rho(E)e^{-\beta E} E^2 dE
\\
&&-
\left(
\frac{1}{Z(\beta)}
\int_{-\infty}^\infty\rho(E)e^{-\beta E} E dE
\right)^2
\Bigg]
\nonumber \ .
\end{eqnarray}
For the susceptibility we employ the $\op{S}^z$-symmetry of Heisenberg systems 
and decompose the density into contributions from all orthogonal subspaces
with total magnetic quantum number $M$, i.e.
\begin{eqnarray}\label{E-2-2-C}
\frac{\chi(\beta)}{(g \mu_B)^2}=
\beta
\Bigg[
&&
\frac{1}{Z(\beta)}
\sum_M M^2
\int_{-\infty}^\infty\rho(E,M)e^{-\beta E} dE
\\
&&-
\left(
\frac{1}{Z(\beta)}
\sum_M M
\int_{-\infty}^\infty\rho(E,M)e^{-\beta E} dE
\right)^2
\Bigg]
\nonumber \ .
\end{eqnarray}
We further calculate only contributions for $M\geq 0$, since the respective
contributions for negative $M$ are degenerate and can be added accordingly.

The idea of the Chebyshev algorithm is to expand the microcanonical density of states 
$\rho(E)$ in terms of Chebyshev polynomials and then approximate the integral 
\fmref{E-2-2} by Gauss-Chebyshev integration. We would like to state already 
at this stage that some accuracy problems shown later in this article
arise if the approximated density of states 
does not behave like a proper density, e.g,\ if it becomes negative.

Since the Chebyshev polynomials are restricted to the interval $[-1, 1]$, 
a variable transformation of the Chebyshev polynomials to arbitrary intervals 
must be introduced as in \cite[Sec. 1.3.2]{mason2002chebyshev}.
The transformation of the Hamiltonian results in
\begin{align}\label{E-2-3}
\op H' := \frac{1}{m}(\op H - c\cdot\op 1)
\end{align}
with 
\begin{align}\label{E-2-4}
m=
\frac{E_{\mbox{\scriptsize max}} - E_{\mbox{\scriptsize min}}}{2-\varepsilon},
\qquad 
c=
\frac{E_{\mbox{\scriptsize max}} + E_{\mbox{\scriptsize min}}}{2}
\end{align}
where, as suggested in \cite{WWA:RMP06}, a parameter $\varepsilon$ is introduced to prevent 
truncation of the approximated delta peaks corresponding to the extremal eigenvalues.
The original energy interval is thus scaled to the interval $[-1+\tfrac{\varepsilon}{2}, 1-\tfrac{\varepsilon}{2}]$.\\
The corresponding scaled density of states $\overline{\rho}(x)$ is then expanded in terms of
Chebyshev polynomials 
$C_n(x)$:
\begin{align}\label{E-2-5}
	\overline{\rho}(x) \approx \frac{1}{\pi\sqrt{1-x^2}}\left[\mu_0 + 2\sum_{n=1}^{N_{\text{\scriptsize deg}}}\ \mu_n C_n(x)\right]\ .
\end{align}
It can be shown that the coefficients of the expansion are given by the traces  
\begin{align}\label{E-2-6}
	\mu_n = \Tr\big[ C_n(\op H')\big]\ .
\end{align}
These traces are approximated using the typicality approach, i.e. 
\begin{align}\label{E-2-7}
	\mu_n\approx \frac{\mbox{dim}\mathcal(H)}{R}\sum_{r=1}^R \frac{\bra{r} C_n(\op H') \ket{r}}{\braket{r}{r}} = \Theta_n(R) \ ,
\end{align} 
with
\begin{eqnarray} \label{eq:te}
\label{E-1-2}
\ket{r}
&=&
\sum_{\nu}\;
r_{\nu} \ket{\nu}
\end{eqnarray}
being a random vector with Gaussian distributed components $r_{\nu}$ with
respect to a chosen orthonormal basis $\{\ket{\nu}\}$.
The relative error of an estimate $\Theta_n(R)$ is proportional 
to $1/\sqrt{R\  \mbox{dim}(\mathcal H)}$ as shown in e.g.\ \cite{WWA:RMP06}, where $R$ is the number 
of random vectors and $\mbox{dim}(\mathcal H)$ the dimension of the Hilbert space.
At this point it should be noted that these traces can also be evaluated to numerical accuracy using a complete basis. 
This possibility will be used to distinguish statistical and systematic deviations later on.  

Due to the finite order of the expansion, so-called Gibbs' oscillations can occur which cause the approximated 
density of the states to have negative values. If one wants to obtain a ``physical" 
representation of the density, i.e.\ without negative values, 
one can modify the coefficients $\mu_n$ by a kernel \cite{WWA:RMP06}. 
The kernel fixes this problem at the cost of introducing a systematic error 
which vanishes for $N_{\text{\scriptsize deg}}\rightarrow\infty$.
In this paper, we restrict our discussion to the use of the Jackson kernel
whose coefficients read
\begin{align}\label{E-2-8}
    g_n&=\frac{1}{\Ndeg+1}\left[(\Ndeg-n+1)\cos\left(\frac{\pi n}{\Ndeg+1}\right)\right.\notag 
    \\\ \notag\\
    &+
    \left.\sin\left(\frac{\pi n}{\Ndeg+1}\right)\cot\left(\frac{\pi}{\Ndeg+1}\right){\Ndeg+1}\right]
    \ .
\end{align} 
In figure captions or legends we will write $g_n=$JK when the kernel is applied, otherwise $g_n=1$.

For an arbitrary function $f(x)$ the Gauss-Chebyshev integration gives rise to the approximation
\begin{align}\label{quad}
   \int_{-1}^{1}\frac{f(x)}{\sqrt{1-x^2}}\ dx \approx \frac{\pi}{\tilde{N}}\sum_{k=1}^{\tilde{N}}f(x_k)\ , \quad \text{see}\  \text{\cite{TdM}}\ ,
\end{align}
where the supporting points read
\begin{align}
    x_k=\cos\left(\frac{\pi(k-\tfrac{1}{2})}{\tilde{N}}\right),\quad k=1,\dots,\tilde{N}
    \ .
\end{align}
This approximation is an exact identity if $f(x)$ is a 
polynomial of order $2\tilde{N}-1$ or smaller \cite{mason2002chebyshev}. 
In the case at hand, $f(x)$ has to be chosen as
\begin{align}
    f(x) := \left[g_0\mu_0 + 2\sum_{n=1}^{N_{\text{\scriptsize deg}}}\ g_n\mu_n C_n(x)\right]e^{\beta(m x + c)}
    \ ,
\end{align}
and is thus no polynomial of order smaller than $2\tilde{N} - 1$. 
However, the approximation through Gauss-Chebyshev integration is still a 
good choice for numerical purposes as it can be computed through a discrete cosine-transform (type III).
Deploying the $f(x)$ to \eqref{quad} gives
\begin{align}
Z 
&=
\int_{-\infty}^\infty\rho(E)e^{-\beta E}dE
\\
&=
\int_{-1}^{1}\overline{\rho}(x)e^{\beta(m x + c)}\ dx\\
&\approx
\frac{1}{\tilde{N}}\sum^{\tilde{N}}_{k=1} \gamma_k \ e^{-\beta(m x_k + c)}
\ ,
\end{align}
where the values of the weights $\gamma_k$ read
\begin{align}
    \gamma_k 
    &:= 
    \pi\sqrt{1-x_k^2}\ \overline{\rho}_M(x_k)\\
    &= 
    g_0\mu_0 +
    2\sum_{n=1}^{\Ndeg} g_n\mu_n \cos
    \left(\frac{n \pi(k-\tfrac{1}{2})}{{\tilde{N}}}\right)
    \ .
\end{align}
If one chooses $\tilde{N}\geq\Ndeg$, the sum can be complemented to an upper limit of $\tilde{N}$ with additional $g_n=0$ terms. 
The $\gamma_k$
can then be computed through a discrete cosine-transform (type III) of the coefficients $g_n\mu_n$ 
which allows a faster computation of the sum. The time needed scales with $\tilde{N}\ln\tilde{N}$ instead of $\tilde{N}\Ndeg$ \cite{WWA:RMP06}.

If there are known symmetries the scheme can be performed for each orthogonal subspace $\mathcal{H}_{\Gamma}$ separately. 
The approximated partition function can then be written as

\begin{align}
    Z(\beta) = \frac{1}{\tilde{N}}\sum_{\Gamma}\sum^{\tilde{N}}_{k=1} \gamma^{\Gamma}_k \ e^{-\beta(m_{\Gamma} x_k + c_{\Gamma})}
    \ .
\end{align}
All of the following systems possess $\op{S}^z$ symmetry which implies orthonormal subspaces 
corresponding to the quantum number of the total magnetization $\Gamma=M$. 
In our numerical examples, subspaces with dimension $D<15,000$ are fully diagonalized for larger systems. 
For smaller systems, i.e. those containing only subspaces of dimension $D<15,000$, 
only subspaces with dimension $D<1,000$ are fully diagonalized. In a real application one would of course
diagonalize all subspaces numerically exactly where this is possible.

To summarize, the Chebyshev method depends on several parameters that can have an effect 
on the accuracy of the results. 
These are the order of the expansion $\Ndeg$, the scaling parameter $\varepsilon$, 
the number of random vectors $R$, the use of a kernel $g_n$, 
and the number of supporting points $\tilde{N}$.

\section{Numerical results}
\label{sec-3}

The Chebyshev algorithm uses random vectors for trace estimation, 
compare \eqref{E-2-7},
therefore the results are expected to exhibit a statistical distribution. 
To assess this statistical behavior, we perform two kinds of studies:
(A) We investigate a thermodynamic observable as a function of the number $R$
of random vectors used for the trace estimator \fmref{E-2-7}, and 
(B) we study the variance among 
$P$ realizations per fixed parameter set (for some of the cases presented below).

By considering each realization $O_i(\beta)$ as a random measurement, 
a mean $\overline{O}(\beta)$ and a variance $\delta O(\beta)^2$ can be defined:   
\begin{align}\label{E-3-1}
    \overline{O}(\beta) &= \frac{1}{P}\sum_{i=0}^{P-1}O_i(\beta)\ ,\\
    \delta O(\beta)^2  &= \overline{O^2}(\beta)-\overline{O}(\beta)^{\, 2}\ .\label{E-3-2}
\end{align}
If additionally an exact result $O_{E}$ is known, the systematic deviation
\begin{align}\label{E-3-3}
    \Delta O(\beta) = \vert\overline{O}(\beta)-O_{\text{\scriptsize E}}(\beta)\vert
\end{align}
can be defined.

\begin{figure}[ht!]
    \centering
    \includegraphics[width=0.69\columnwidth]{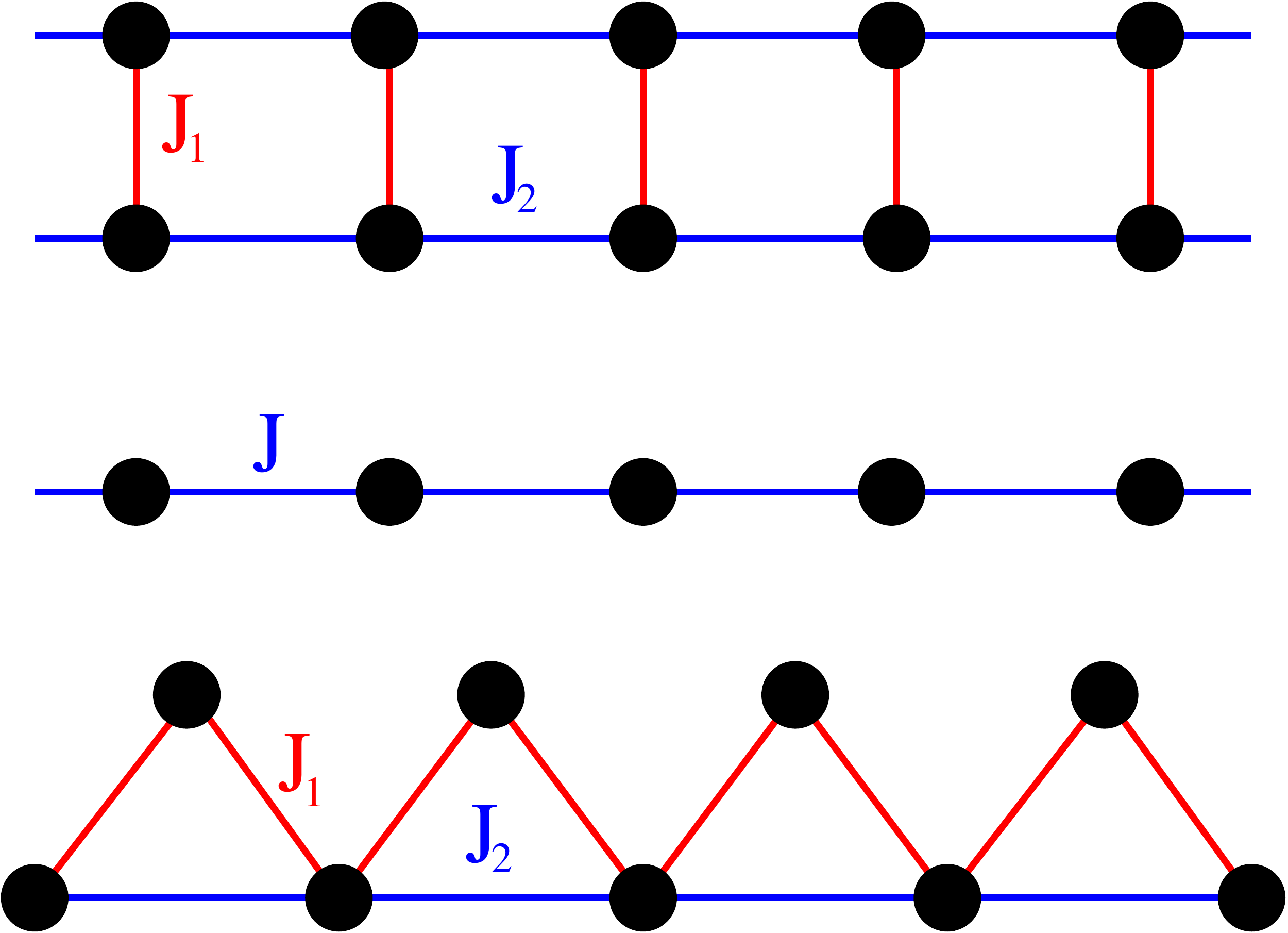}
    \caption{Systems investigated in sec.~\xref{sec-3-1}, from top to bottom: ladder, chain, sawtooth chain. 
    Periodic boundary conditions will be applied.\label{fig-systems}}
 \end{figure} 

As specific quantum spin systems we investigate three archetypical systems
that show fundamentally different behavior at low temperatures, namely
a spin ladder that is gapped in the thermodynamic limit
\cite{DaR:Science96}, 
a spin chain that is gapless in the thermodynamic limit \cite{MiK:LNP04}, 
and a sawtooth chain in the vicinity of a 
quantum-critical point \cite{KDN:PRB14,DKR:PRB18,BML:npjQM18}, 
compare \figref{fig-systems}.

\subsection{Heisenberg ladder}
\label{sec-3-1}
In this subsection, the accuracy of the Chebyshev algorithm is investigated 
using a Heisenberg ladder for various numbers of spins $N$ with spin quantum number $s=1/2$ 
and periodic boundary conditions. The Hamiltonian reads
\begin{eqnarray}
\label{E-3-1-1}
\op{H}
&=&
J_1
\sum_{i}\;
\op{\vec{s}}_{i,1} \cdot \op{\vec{s}}_{i, 2}
+
J_2
\sum_{i,j}\;
\op{\vec{s}}_{i,j} \cdot \op{\vec{s}}_{i+1,j}
\\
&&
+
g \mu_B\, B\,
\sum_{i,j}\;
\op{s}^z_{i,j}
\nonumber
\ .
\end{eqnarray}
where the first subscript $i\in\{1,\dots,N/2\}$ of the spin operators denotes the rung and the second subscript $j\in\{1,2\}$ denotes the leg of the spin. 
Thus, the exchange interaction $J_1$ connects nearest neighbor spins on rungs, $J_2$ does
the same on legs. Both are chosen to be antiferromagnetic, $J_1=J_2=1$. 
\renewcommand*{\arraystretch}{1.5}
\begin{table}[ht!]
    \centering
    \begin{tabular}{|c|c|c|c|c|}
    \hline
        $\quad R\quad$ & $\quad\Ndeg\quad$ & $\quad\varepsilon\quad$ & $\quad\tilde{N}\quad$ & $\quad g_n \quad$ \\
        \hline
        $200$ & $100$ & $0$ & $100$ & $1$\\
        \hline
    \end{tabular}
    \caption{Standard configuration of parameters used in the following calculations. \label{K0}}
\end{table}
Table~\xref{K0} shows the standard configuration of parameters used in the following.
The order of expansion $\Ndeg$ and the number of random states $R$ 
are chosen for low computation times and sufficiently accurate results. 
Their influences on the accuracy are discussed in \secref{sec-3-1-2-1} and 
\secref{sec-3-1-1}. As the parameter $\varepsilon$ is introduced as a corrective variable, 
its influence will be shown separately in \secref{sec-3-1-2-2} and is omitted for the time being. 
The same argument holds for the kernel $g_n$, shown in \secref{sec-3-1-2-4}. 
To make use of the discrete cosine-transform (type III), the number of points of integration $\tilde{N}$ 
has to be greater than or equal to $\Ndeg$. The equality is chosen as a starting point.

\subsubsection{Statistical deviations}
\label{sec-3-1-1}
\begin{figure}[ht!]
    \centering
    \includegraphics[width=0.69\columnwidth]{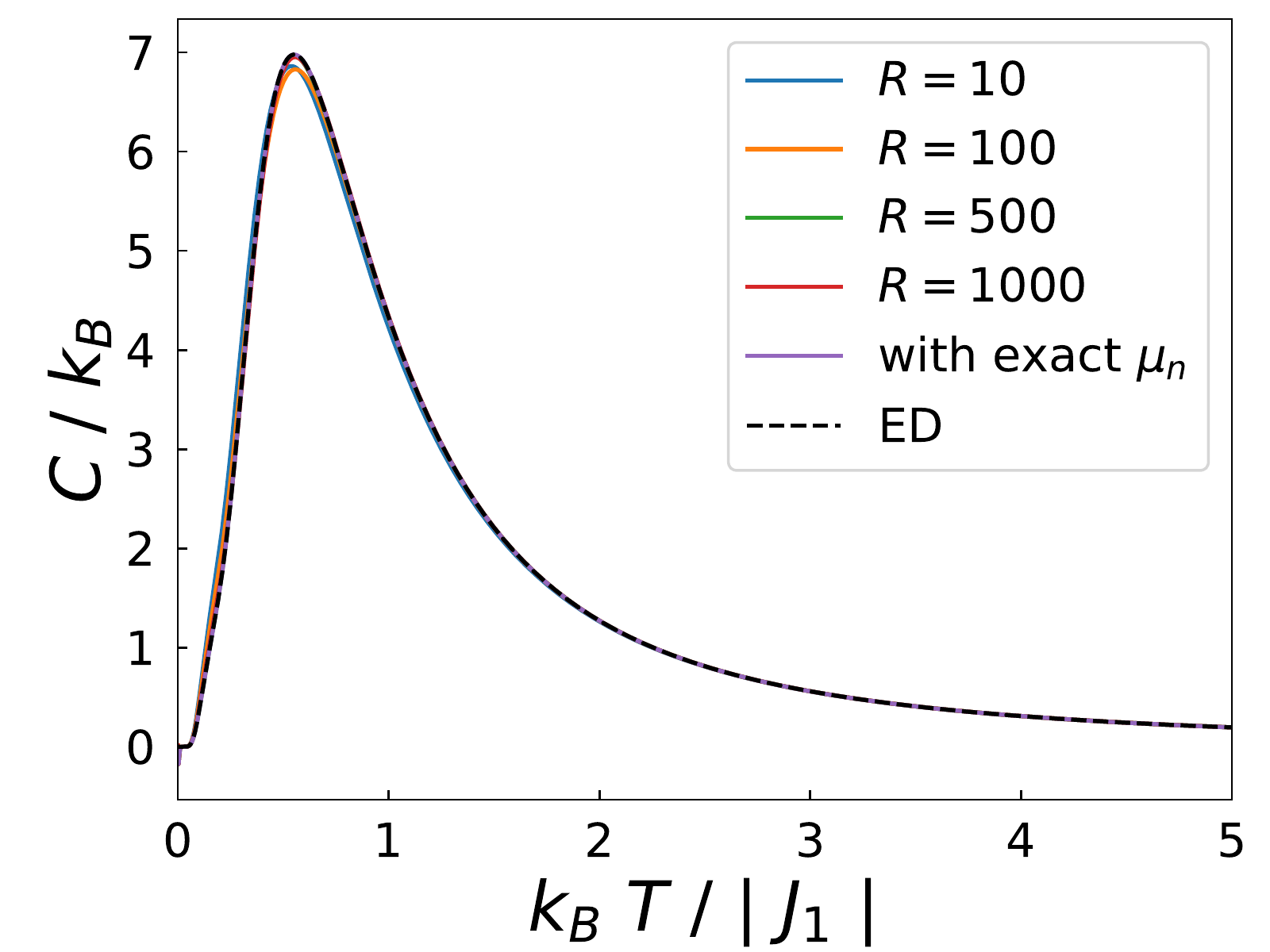}
    \includegraphics[width=0.69\columnwidth]{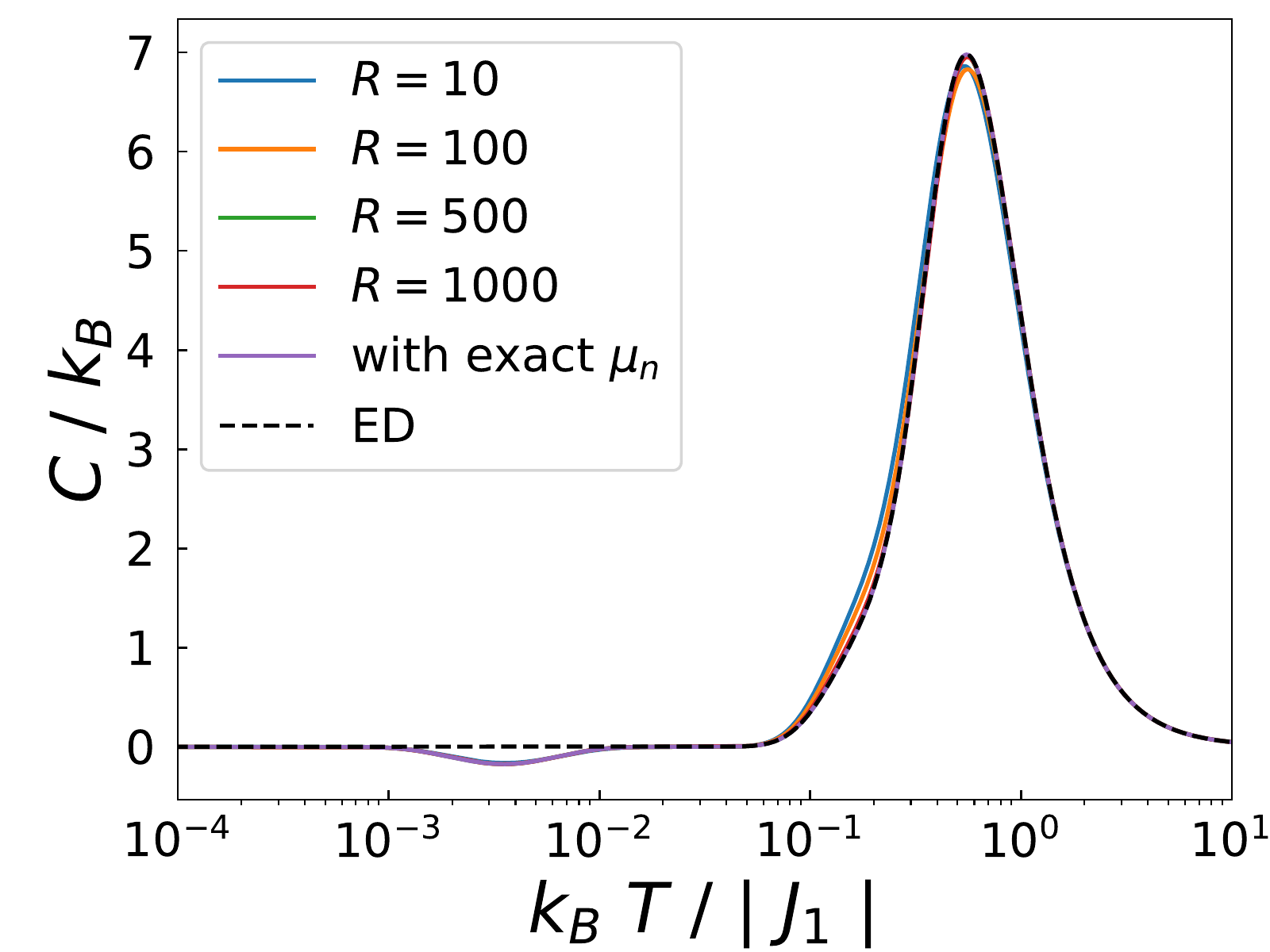}
    \caption{The heat capacity of the Heisenberg ladder with $N=16$ spins $s=1/2$ 
    at zero field in linear and logarithmic plots computed using the Chebyshev algorithm 
    in the standard parameter configuration, see \tabref{K0}, 
    for various values of $R$ (colored curves) and with exact diagonalization (ED). The purple curve shows a result of the Chebyshev algorithm where the calculation of the traces $\mu_n$ is done using a complete basis. 
    \label{cAvR}}
 \end{figure}

\begin{figure}[ht!]
    \centering
    \includegraphics[width=0.69\columnwidth]{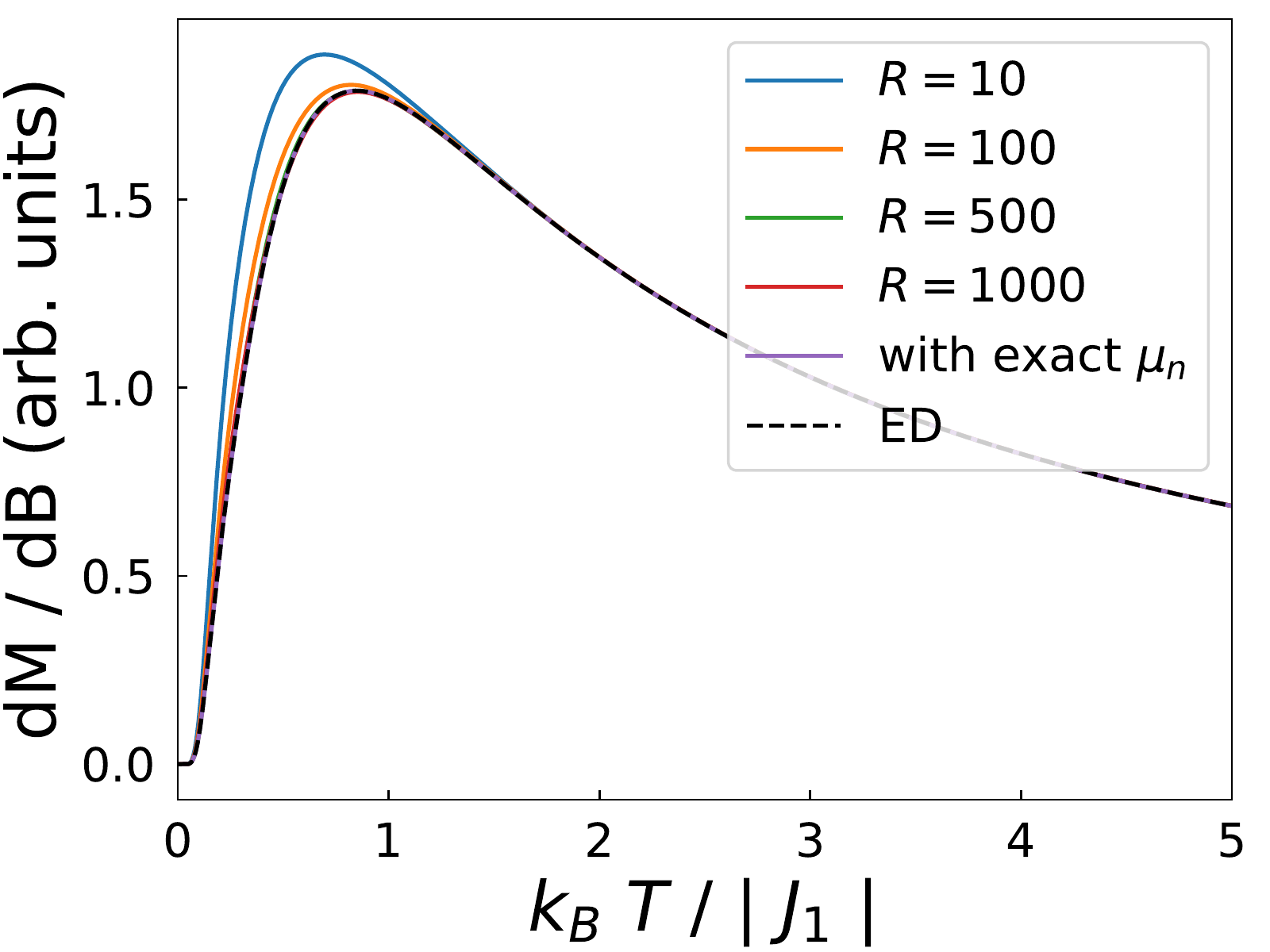}
    \includegraphics[width=0.69\columnwidth]{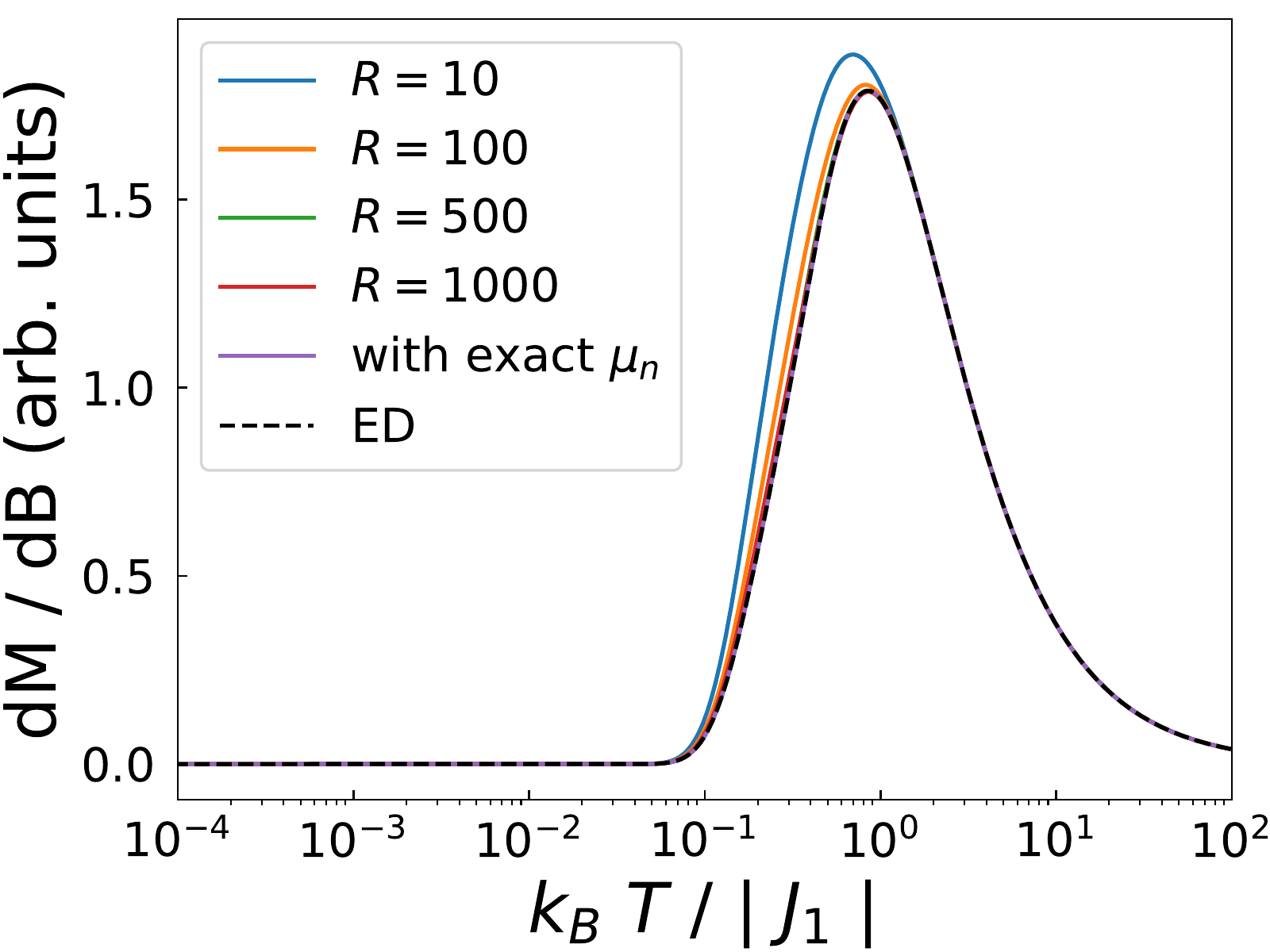}
    \caption{The differential magnetic susceptibility of the Heisenberg ladder with $N=16$ spins $s=1/2$ 
    at zero field in linear and logarithmic plots computed using the Chebyshev algorithm in the standard parameter configuration, see \tabref{K0}, for various values of $R$ (colored curves) and  with exact diagonalization (ED). The purple curve shows a result of the Chebyshev algorithm where the calculation of the traces $\mu_n$ is done using a complete basis. \label{xAvR}}
 \end{figure}

Since the approximation of the traces by using random vectors is the cause of the statistical variations in the result, 
the number of random vectors $R$ is varied to investigate the latter. 
The values given in \tabref{K0} are used as a standard configuration.
In \figref{cAvR} and \ref{xAvR}, the heat capacity and susceptibility for various values of $R$ are plotted next to the result determined by exact diagonalization. 
In addition, a result determined by the Chebyshev algorithm is shown where the traces $\mu_n$ are computed numerically exactly instead of approximating them using random vectors.

One can see that both the heat capacity (\figref{cAvR}) as well as the differential susceptibility (\figref{xAvR})
match the respective curve derived from exact diagonalization very well for $T/\vert J_1 \vert > 10^{-2}$. However, 
for $R=10$ a noticeable deviation in the maximum of the main peak of both observables can be seen. 
Additionally,  all curves of the heat capacity show a ``ghost dip" at low temperatures for 
$T/\vert J_{i} \vert\approx 10^{-3} -10^{-2} $. 
Nevertheless, for most purposes the achieved accuracy for the standard parameter configuration and $R>100$ 
is more than sufficient at higher temperatures. 

It is noticeable that in the region of the ``ghost dip", all approximate curves 
deviate from the exact solution independently of $R$, suggesting  a small statistical 
but significant systematic error. This is confirmed by the fact that the curve 
determined with numerically exact traces shows this deviation as well.

\subsubsection{Systematic deviations}
\label{sec-3-1-2}
Next, we discuss how tuning the parameters $\Ndeg, \varepsilon, \tilde{N}$, and $g_n$ affects systematic deviations.
This is mostly done by observing the behavior of the``ghost dip" of the heat capacity under variation of each parameter. 

\begin{figure}[h!]
 \centering
 \includegraphics[width=0.69\columnwidth]{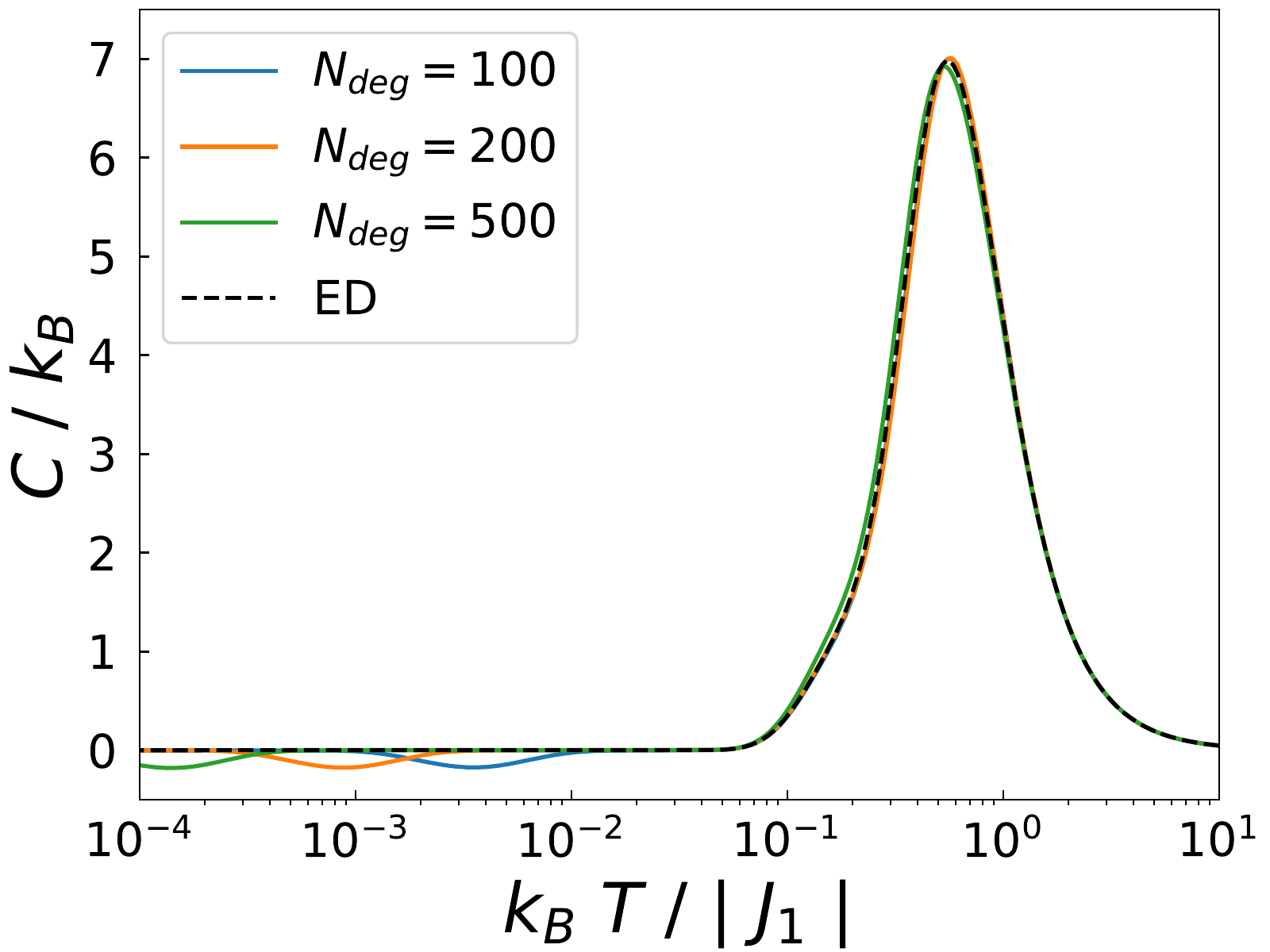}
 \caption{The heat capacity $C / k_B$ of the Heisenberg ladder with $N=16$ 
 spins $s=1/2$ computed using the Chebyshev method in the standard parameter configuration, see \tabref{K0}, 
 for various $\Ndeg$ (colored curves) compared to exact diagonalization (ED).\label{C_Ndeg}}
 \end{figure}

\paragraph{The order of expansion}
\label{sec-3-1-2-1}
can be increased to push the systematic deviations, i.e. the ``ghost dip", to lower temperatures. 
This is demonstrated in \figref{C_Ndeg}. Computation time increases linearly with $\Ndeg$.              


\begin{figure}[h!]
 \centering
 \includegraphics[width=0.69\columnwidth]{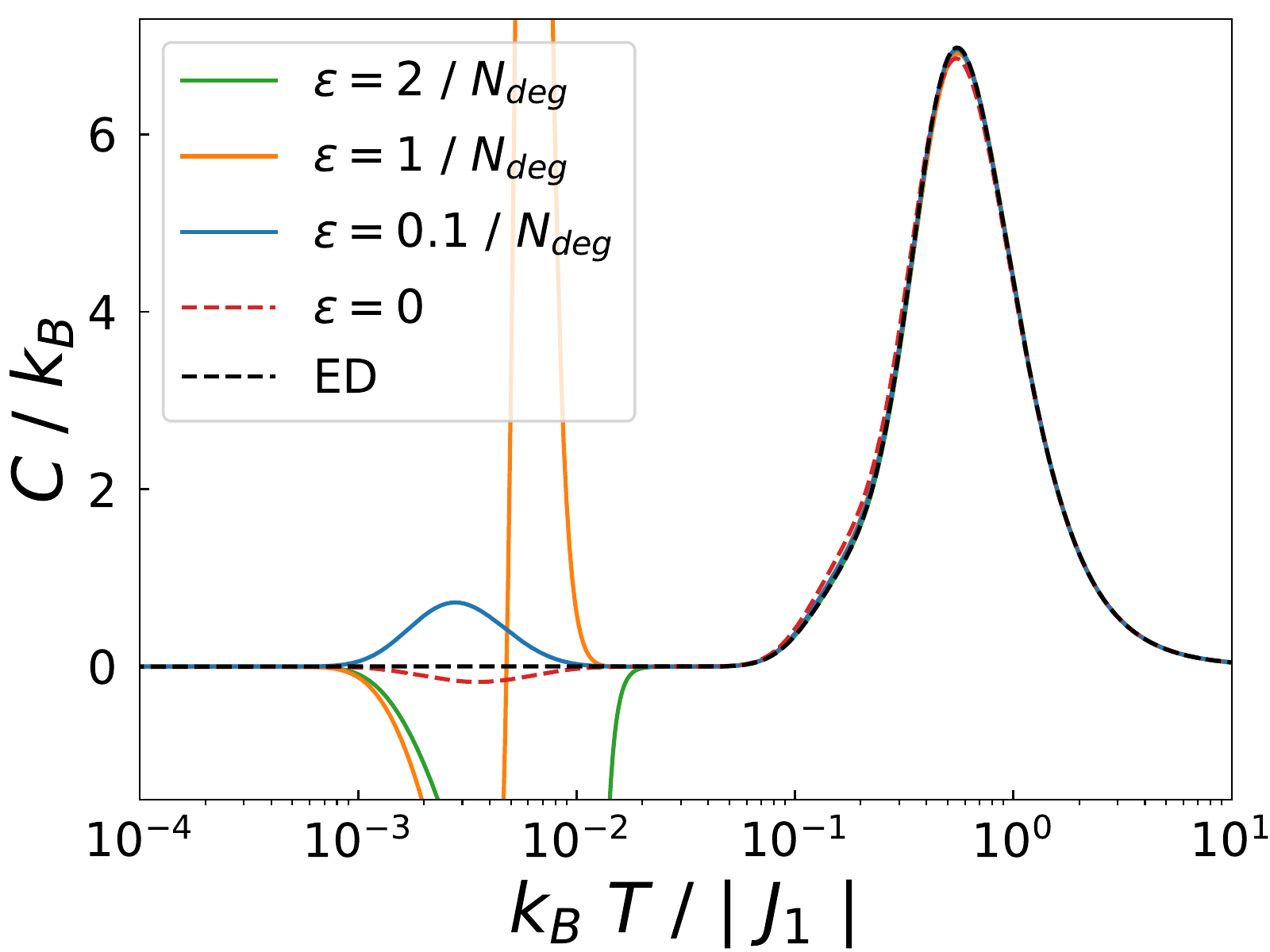}
 \caption{The heat capacity $C / k_B$ of the Heisenberg ladder 
 with $N=16$ spins $s=1/2$ computed using the Chebyshev algorithm in the standard parameter configuration, see \tabref{K0}, 
 for various values of $\varepsilon$ (colored curves) compared to 
 exact diagonalization (ED).\label{C_eps}}
 \end{figure}

\begin{figure}[h!]
 \centering
 \includegraphics[width=0.69\columnwidth]{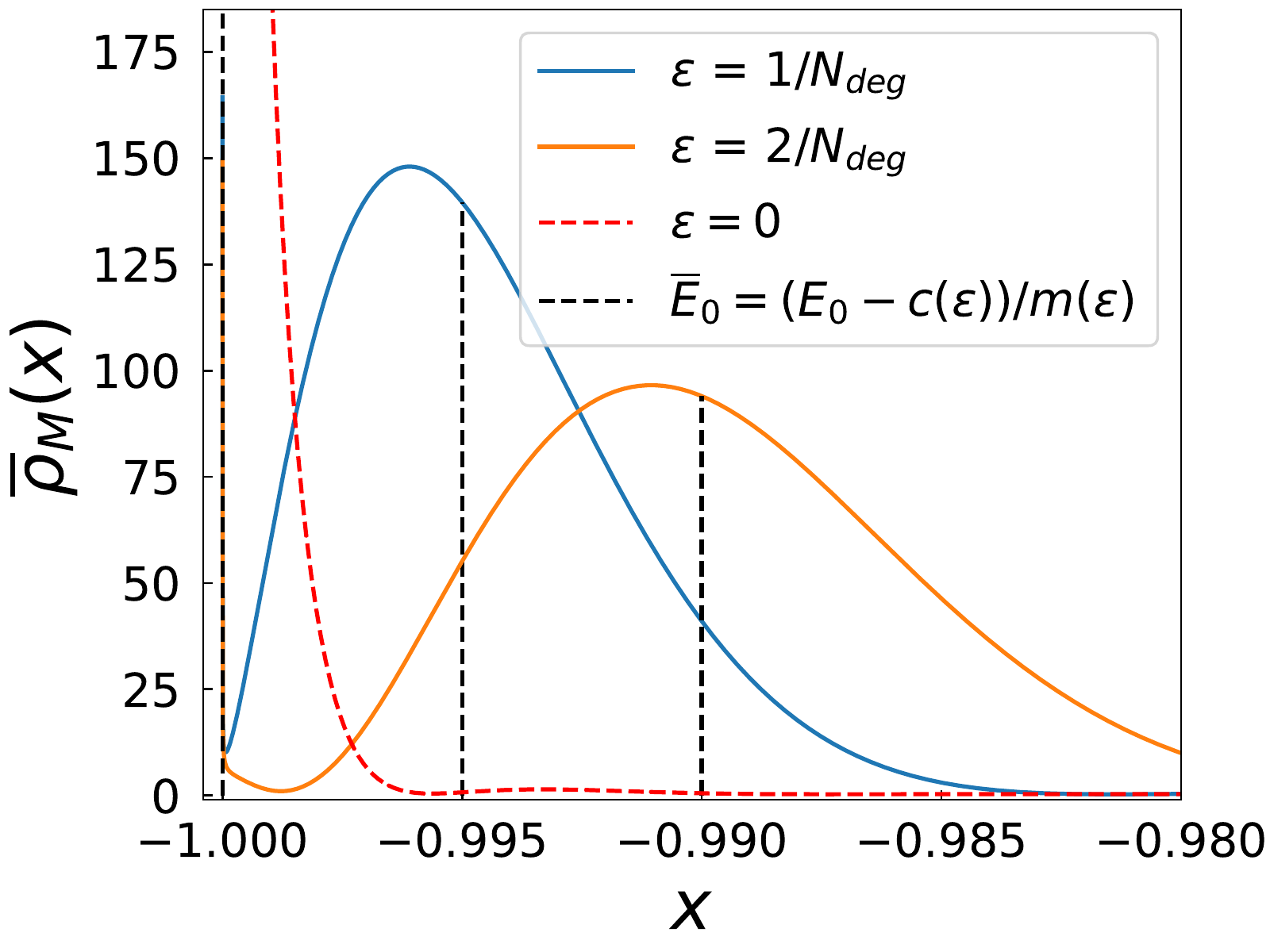}
 \caption{Detail of the scaled density of states $\rho_M(x)$ of the Heisenberg ladder with $N=16$ spins $s=1/2$ on the subspace with $M=0$ computed using the Chebyshev algorithm in the standard parameter configuration, see \tabref{K0}, and the scaled groundstate eigenvalue
 obtained by the Lanczos algorithm for various values of $\varepsilon$.\label{rho_eps}}
 \end{figure}

\paragraph{The scaling parameter $\varepsilon$}
\label{sec-3-1-2-2} seems to have a negative rather than the intended positive effect on the 
heat capacity as shown in \figref{C_eps}. The best result is achieved for $\varepsilon = 0$. 
Conversely, when considering the results for the density of states of the largest subspace with $M=0$ 
(see \figref{rho_eps}), one can see that the parameter $\varepsilon$ has the intended effect
to prevent the peaks of the lowest and highest eigenvalues to be cut off. 
There are also situations where a $\varepsilon\approx 10^{-6}$ seemingly decreases 
the depth of the dip compared to $\varepsilon=0$. 
Anyhow, such small values of $\varepsilon$ are not sufficient to prevent 
the cut-off of the density of states, 
and the improvement was not significant when compared to statistical deviations.

\begin{figure}[h!]
 \centering 
  \includegraphics[width=0.69\columnwidth]{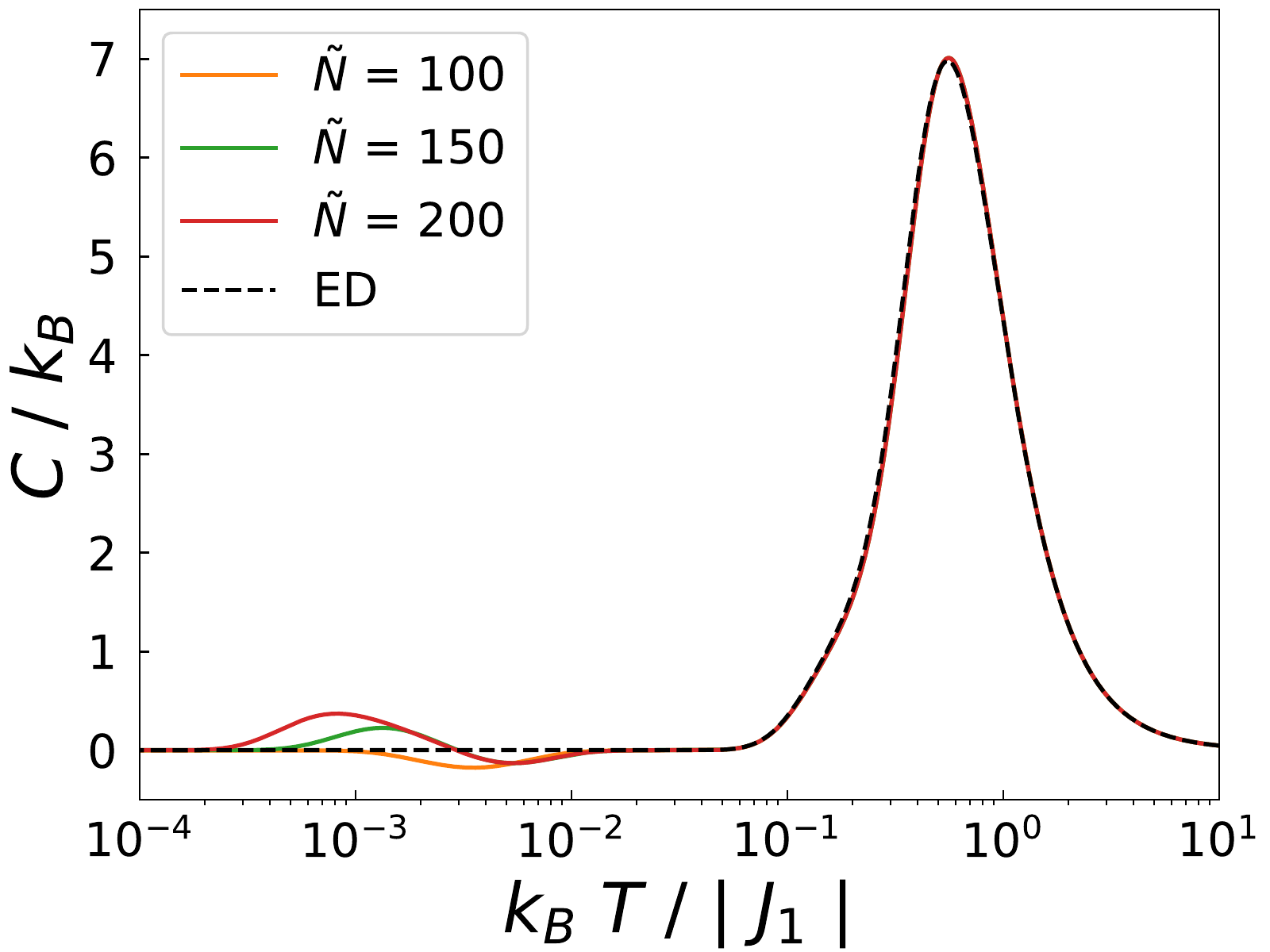}
 \caption{The heat capacity $C / k_B$ of the Heisenberg ladder with $N=16$ spins $s=1/2$ computed using the Chebyshev method in the standard parameter configuration, see \tabref{K0}, for various $\tilde{N}$ (colored) and with exact diagonalization (ED).\label{CNst}}
\end{figure}

\paragraph{The number of supporting points $\tilde{N}$,}
\label{sec-3-1-2-3}
is investigated in \figref{CNst}. It is difficult to give universal recommendations regarding this parameter. 
Our experience is that it should be chosen equal to the order of expansion $\Ndeg$. 
Other systems could show very different results, 
but it is important to note that a good choice of $\tilde{N}$ scales with $\Ndeg$.

\begin{figure}[h!]
 \centering
 \includegraphics[width=0.69\columnwidth]{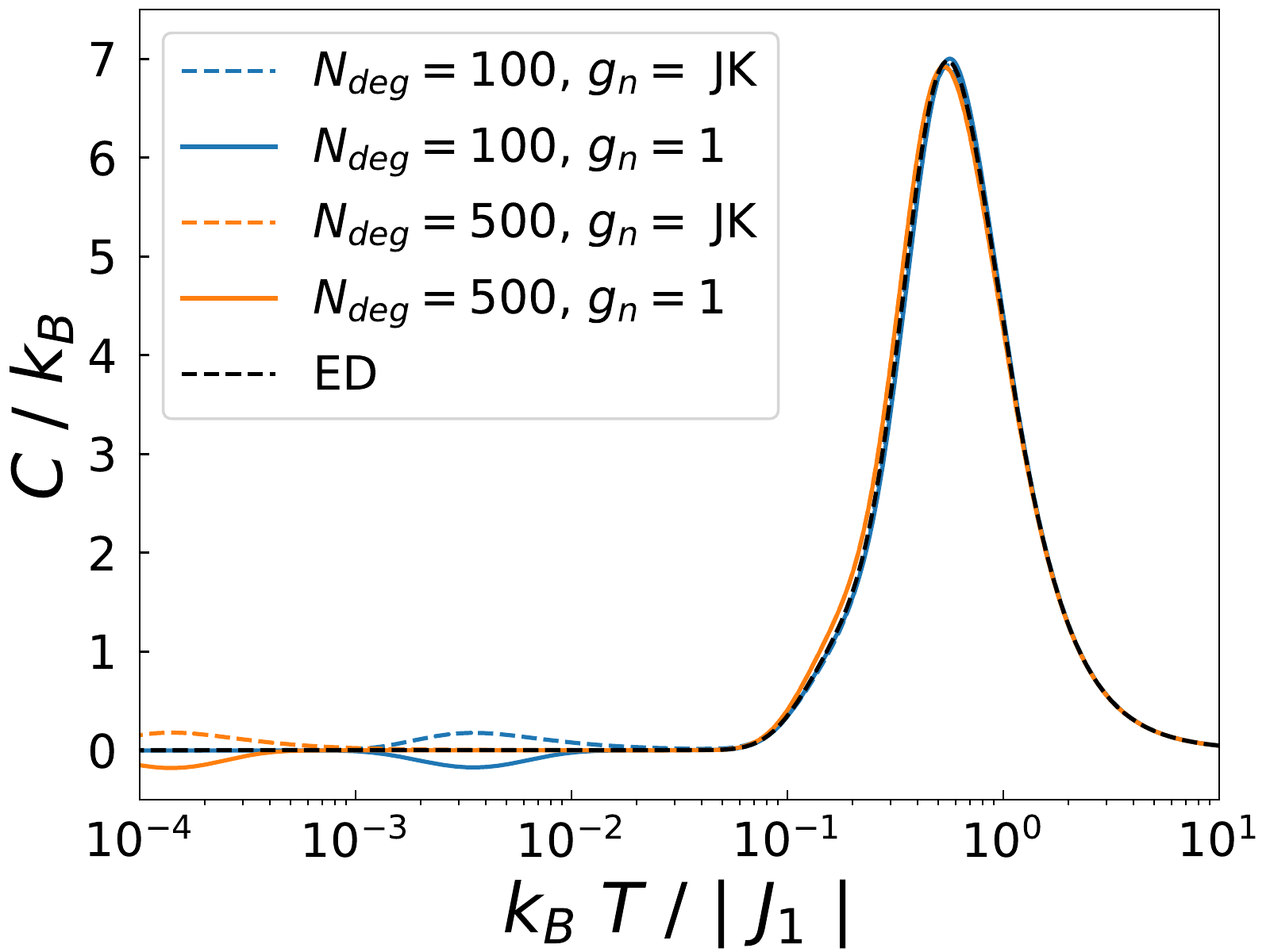}
 \caption{The heat capacity $C / k_B$ of the Heisenberg ladder with $N=16$ spins $s=1/2$  
 computed using the Chebyshev algorithm (colored) in the standard parameter configuration, 
 see \tabref{K0}, for various $\Ndeg$ with and without Jackson kernel compared to
 exact diagonalization (ED).\label{C_Jack}}
 \end{figure}

\paragraph{The smoothing with the Jackson kernel}
\label{sec-3-1-2-4} 
causes the unphysical ghost dip to become a ghost peak, see \figref{C_Jack}, 
but it does not seem to improve the result. It could even produce a negative effect 
as a dip (negative $C$) is more easily identified as an error than a peak.
A good approach could be to always compare both the smoothened and the native result. 
This can be done without great computational effort.  

\begin{figure}[h!]
 \centering
 \includegraphics[width=0.69\columnwidth]{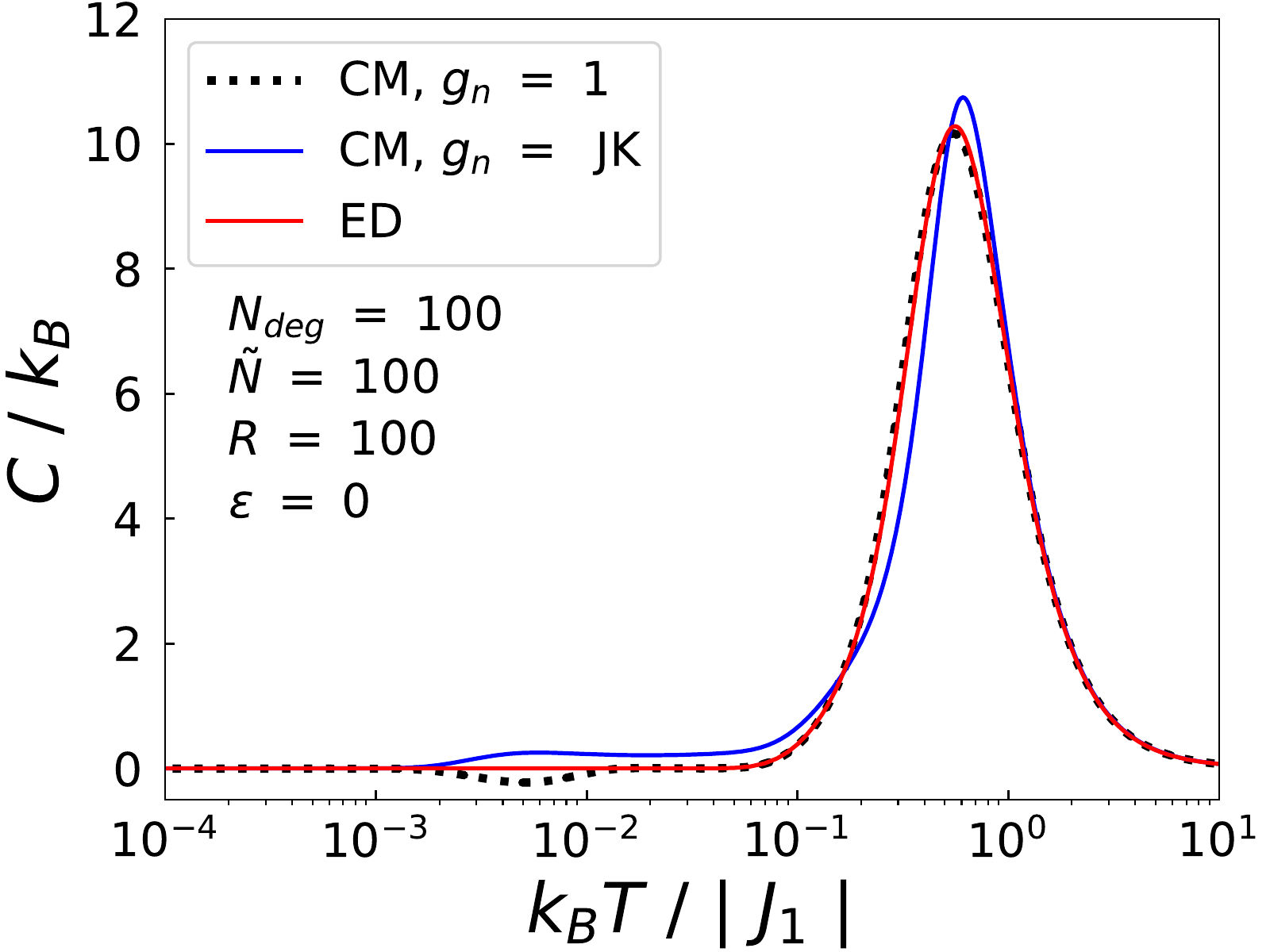}
 \includegraphics[width=0.69\columnwidth]{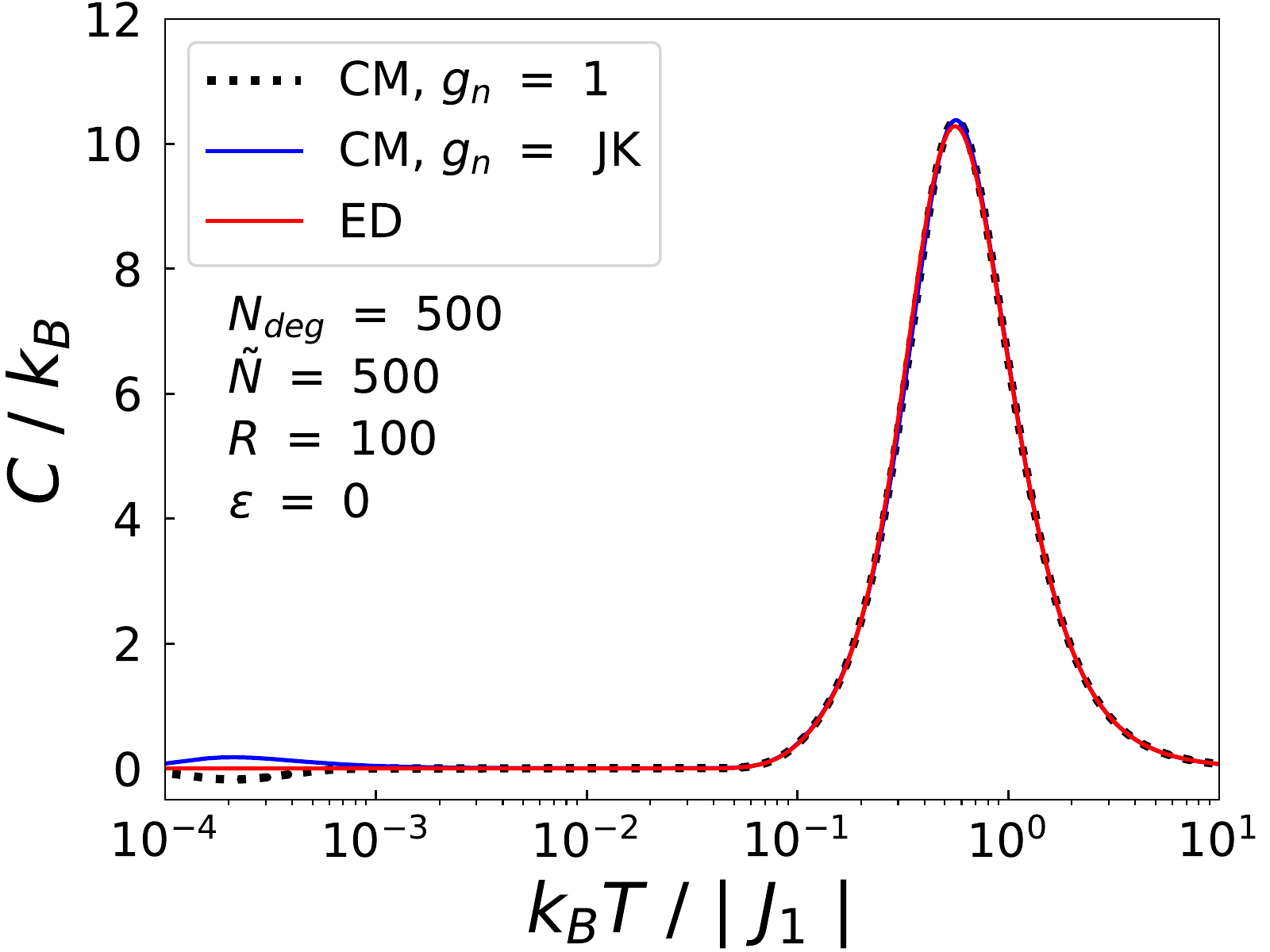}
 \caption{The heat capacity $C / k_B$ of the Heisenberg ladder with $N=24$ spins $s=1/2$ 
 computed using the Chebyshev method (CM) and exact diagonalization (ED). 
 Displayed are the results with and without Jackson kernel (JK) for $\Ndeg = 100$ and $\Ndeg = 500$.\label{Cjb_ladder}}
\end{figure}

For larger systems a kernel can have an even stronger negative effect as is demonstrated
in \figref{Cjb_ladder} for $N=24$ and $s=1/2$. The Jackson kernel is
significantly setting back the convergence of the expansion. 
For an order of $\Ndeg=100$, which produces very accurate approximations without kernel,
the application of the kernel renders the result to become unusable, compare top of \figref{Cjb_ladder}.
One needs to expand the polynomial to an order of $\Ndeg=500$ to counteract the inaccuracy introduced by the kernel,
but even then the result with kernel is still not significantly better than the result without the kernel. 
The result without the kernel seems already sufficiently accurate for $k_B T/|J_1| < 10^{-2}$ and $\Ndeg=100$.
 
5
 \begin{figure}[h!]
 \centering
  \includegraphics[width=0.69\columnwidth]{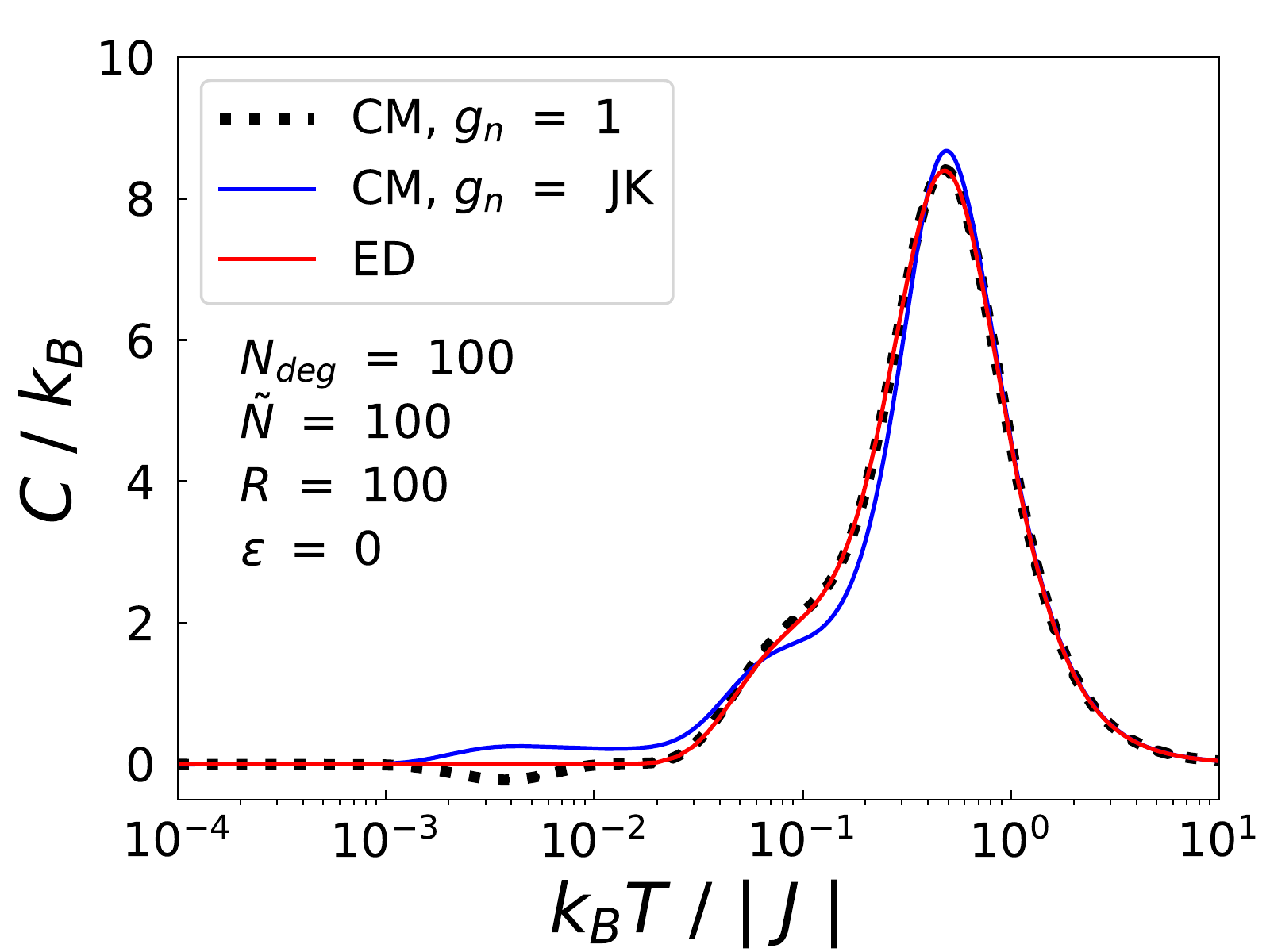}
 \includegraphics[width=0.69\columnwidth]{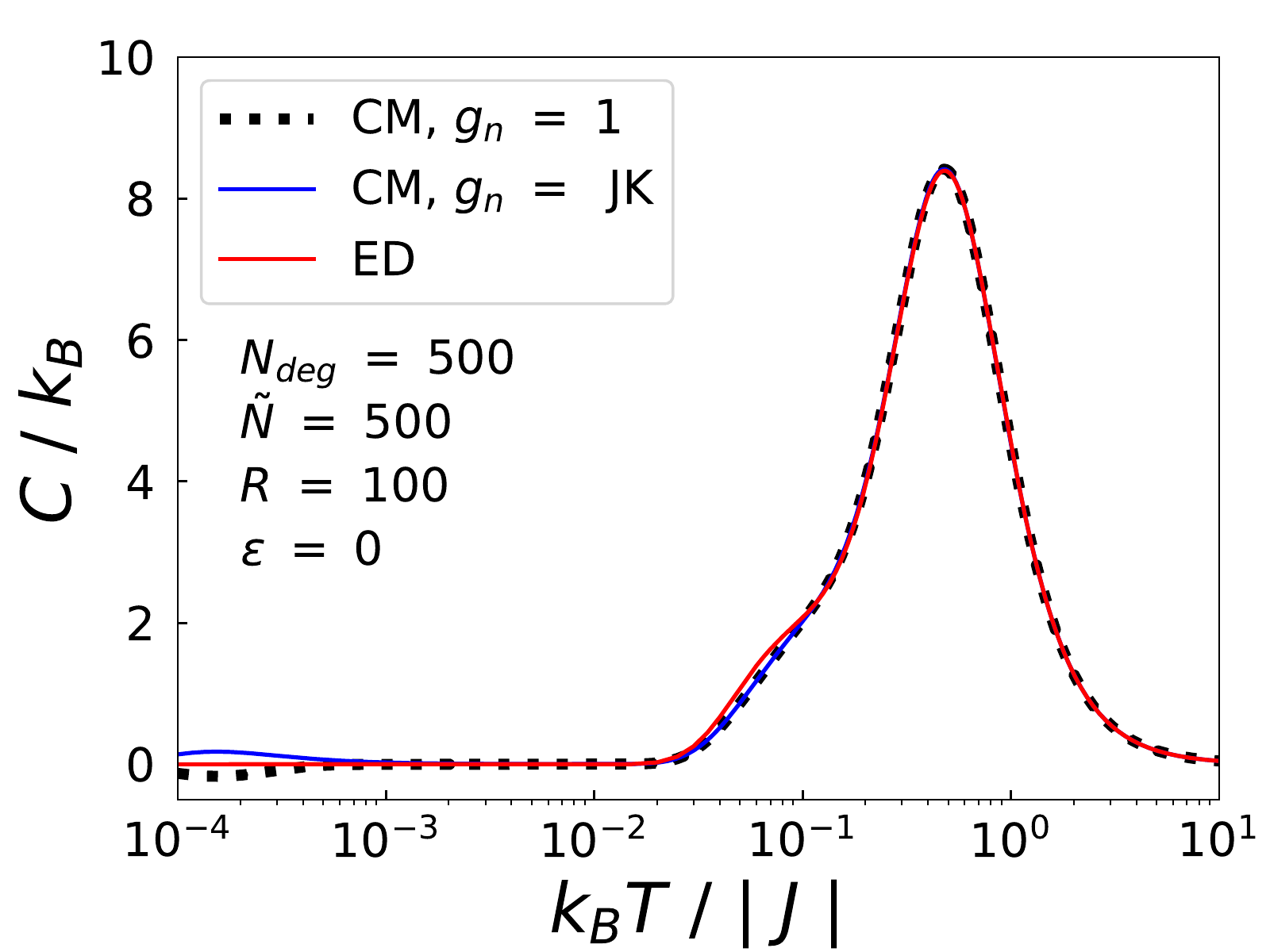}
 \caption{The heat capacity $C / k_B$ of the Heisenberg ring with $N=24$ spins $s=1/2$ computed using the Chebyshev method (CM) and exact diagonalization (ED). Displayed are the results with and without Jackson kernel (JK) for $\Ndeg = 100$ and $\Ndeg = 500$.\label{C_j_ring24}}
 \end{figure}

\begin{figure}[ht!]
 \centering
  \includegraphics[width=0.69\columnwidth]{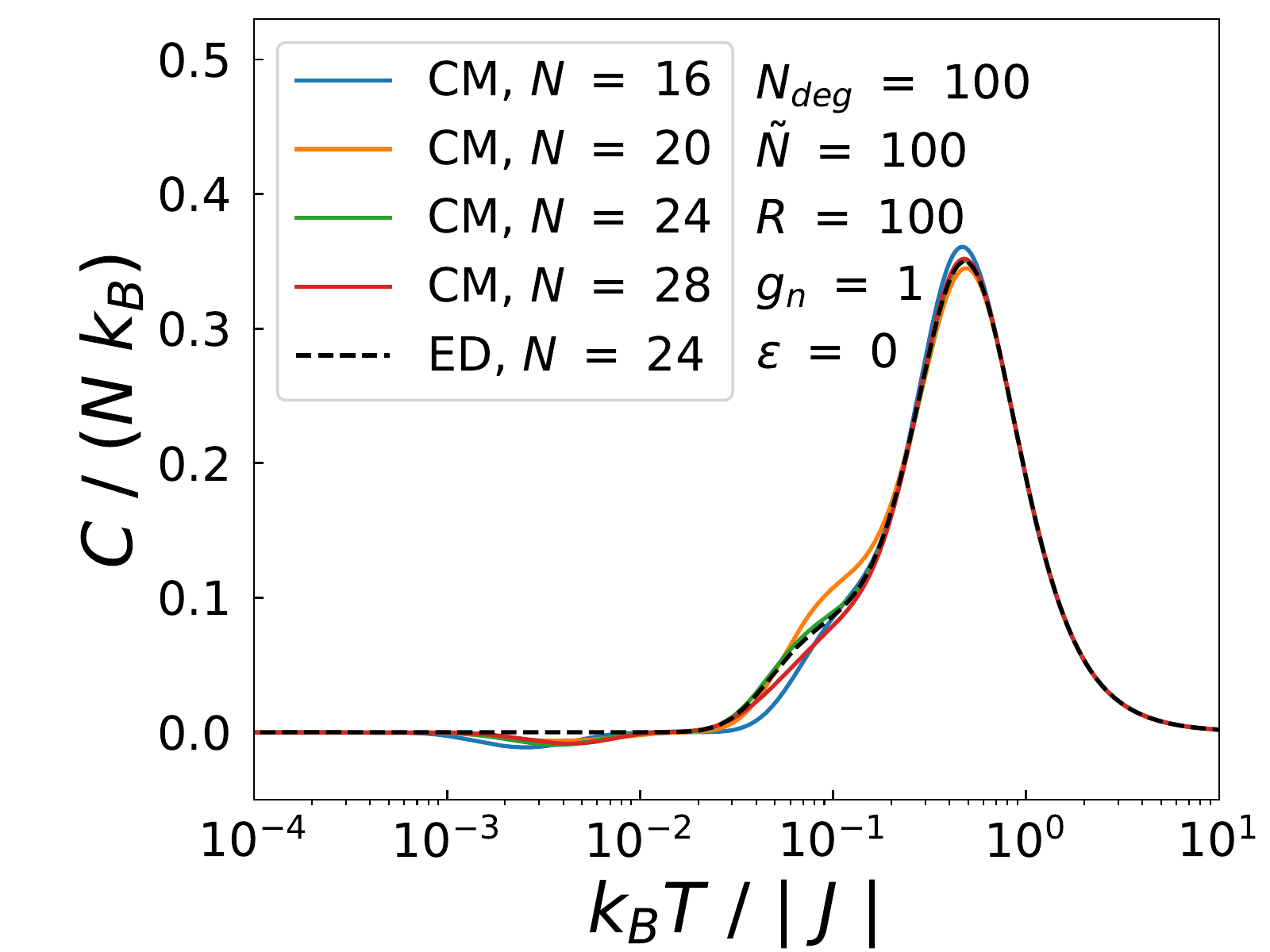}
   \includegraphics[width=0.69\columnwidth]{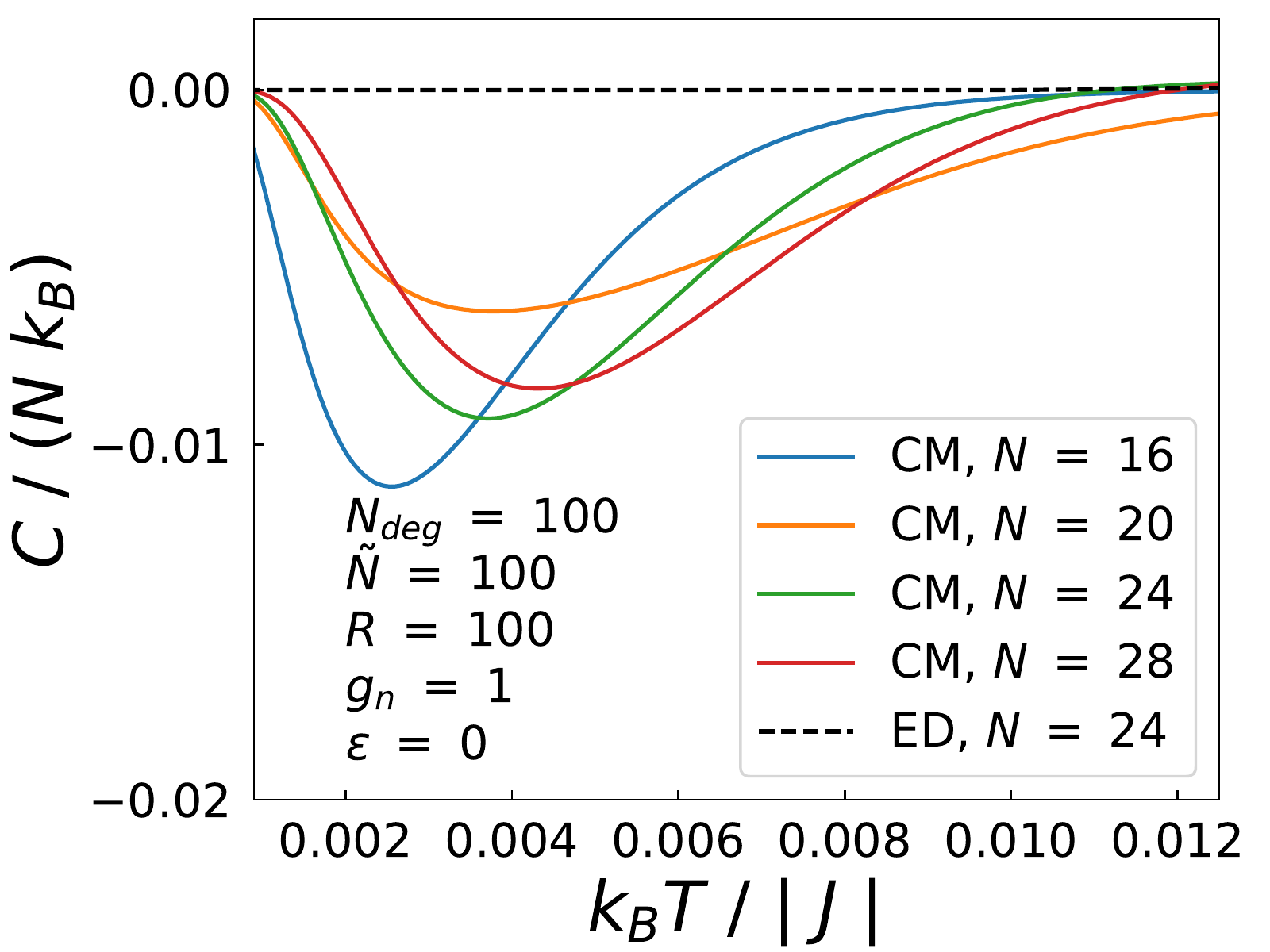}
 \caption{The heat capacity per site $C / (N k_B)$ of the Heisenberg ring with $s=1/2$ computed using the Chebyshev method (CM) for $N = 16,\ 20,\ 24,\ 28$ and exact diagonalization (ED) for $N = 24$.\label{C_FS_ring}}
 \end{figure}

\subsection{Heisenberg ring}

Since the Heisenberg ladder is a gapped system, i.e a spin system with a non-zero excitation energy 
between the groundstate and the first excited state in the thermodynamic limit, we would also like to investigate
a system that is gapless in the thermodynamic limit.
The behavior of thermodynamic functions of the system at low temperatures highly depends 
on this excitation energy. One could argue that the deviation shown in the previous section 
are due to this dependency. 
Hence, in this section we will discuss an antiferromagnetic Heisenberg ring with 
$s=1/2$ for which the excitation energy vanishes in the thermodynamic limit.

In \figref{C_j_ring24}  the results for the Heisenberg ring with $N=24$ spins is displayed. 
One can see that the deviations here are very similar to the ones for the Heisenberg 
ladder with the same system size and choice of parameters, compare \figref{Cjb_ladder}, 
even though here, they occur at slightly lower temperatures.  

To further investigate the influence of the gap's size on the deviations in the results the curves for the heat capacity per site for various numbers of spins are displayed in \figref{C_FS_ring}. While the differences of the results between $k_B T/|J| = 2\cdot 10^{-2}$ and $2\cdot 10^{-1}$ are mostly due to finite-size effects, so even the exact curves would deviate from each other, the differences between  $k_B T/|J| = 10^{-3}$ and $10^{-2}$ on the other hand are due to the method's inability to reproduce the low temperature behavior of the Heisenberg ring for various system sizes. However, there is no definite trend of better results for larger systems recognizable, see the lower graph in \figref{C_FS_ring}. Thus, the inaccuracies of the results do not directly depend on the gap size.

\begin{figure}[ht!]
 \centering
 \includegraphics[width=0.69\columnwidth]{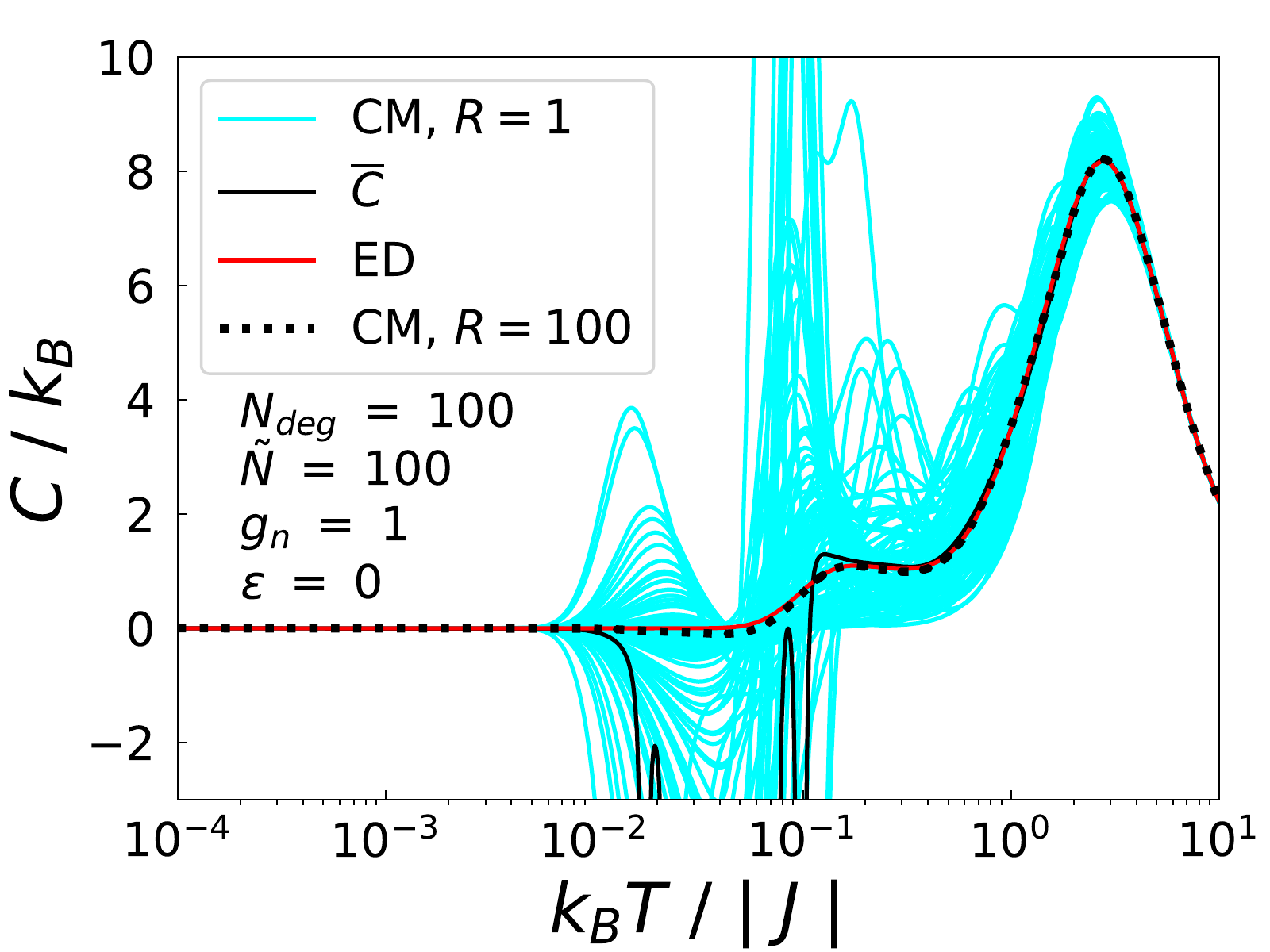}
  \includegraphics[width=0.69\columnwidth]{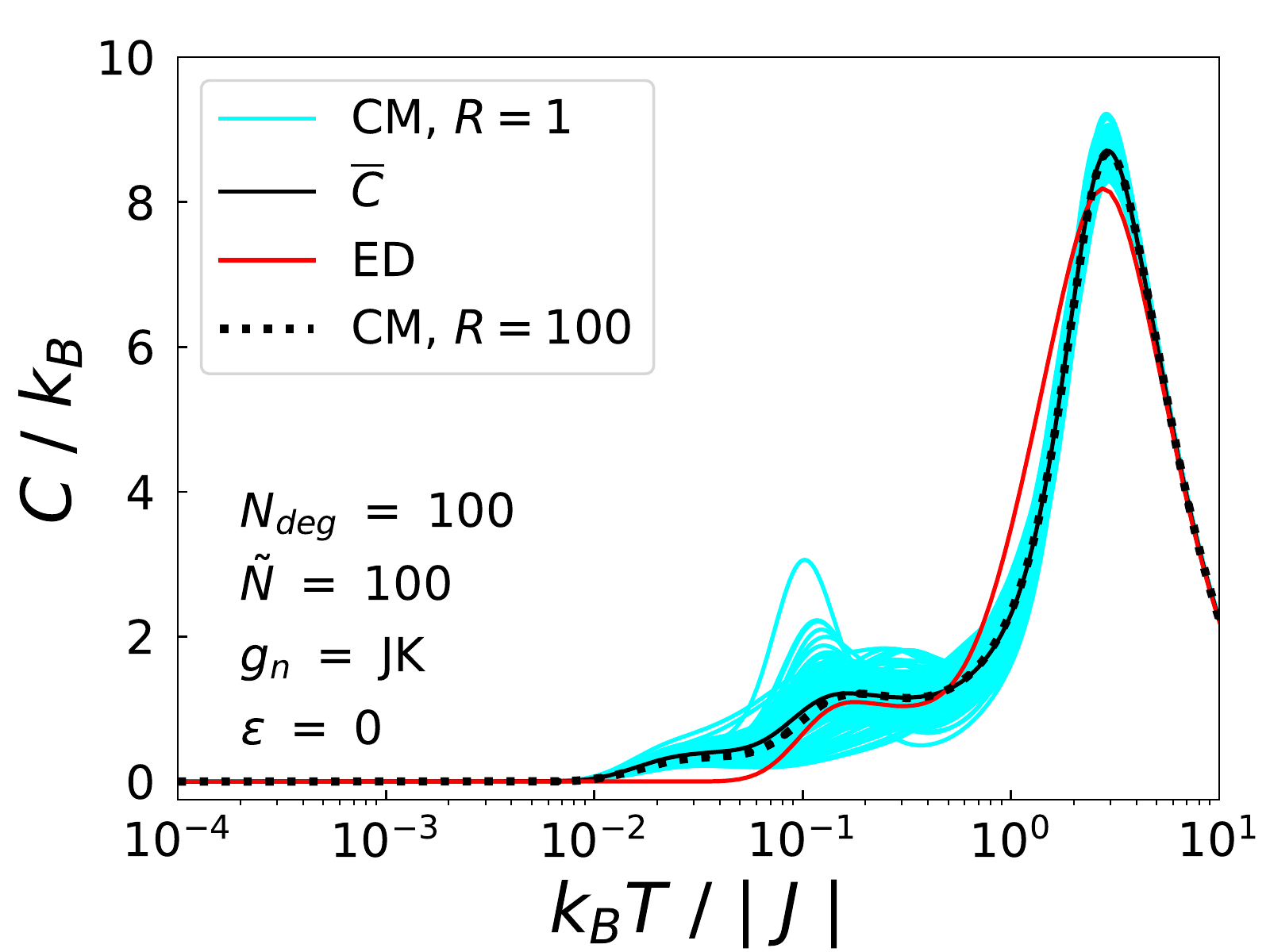}
 \caption{The heat capacity $C / k_B$ of the Heisenberg ring with $N=10$ spins $s=5/2$  computed using the Chebyshev method (CM) and exact diagonalization (ED). 
 Displayed are the results of $P = 100$ realizations with $R=1$ (light blue curves),
 their average (dark solid curve), the exact result (red solid curve), 
 and one realization ($P=1$) with $R=100$ (dotted curve); $\Ndeg=100$.}
 \label{C_R_ring}
 \end{figure}
   
 \begin{figure}[ht!]
 \centering
  \includegraphics[width=0.69\columnwidth]{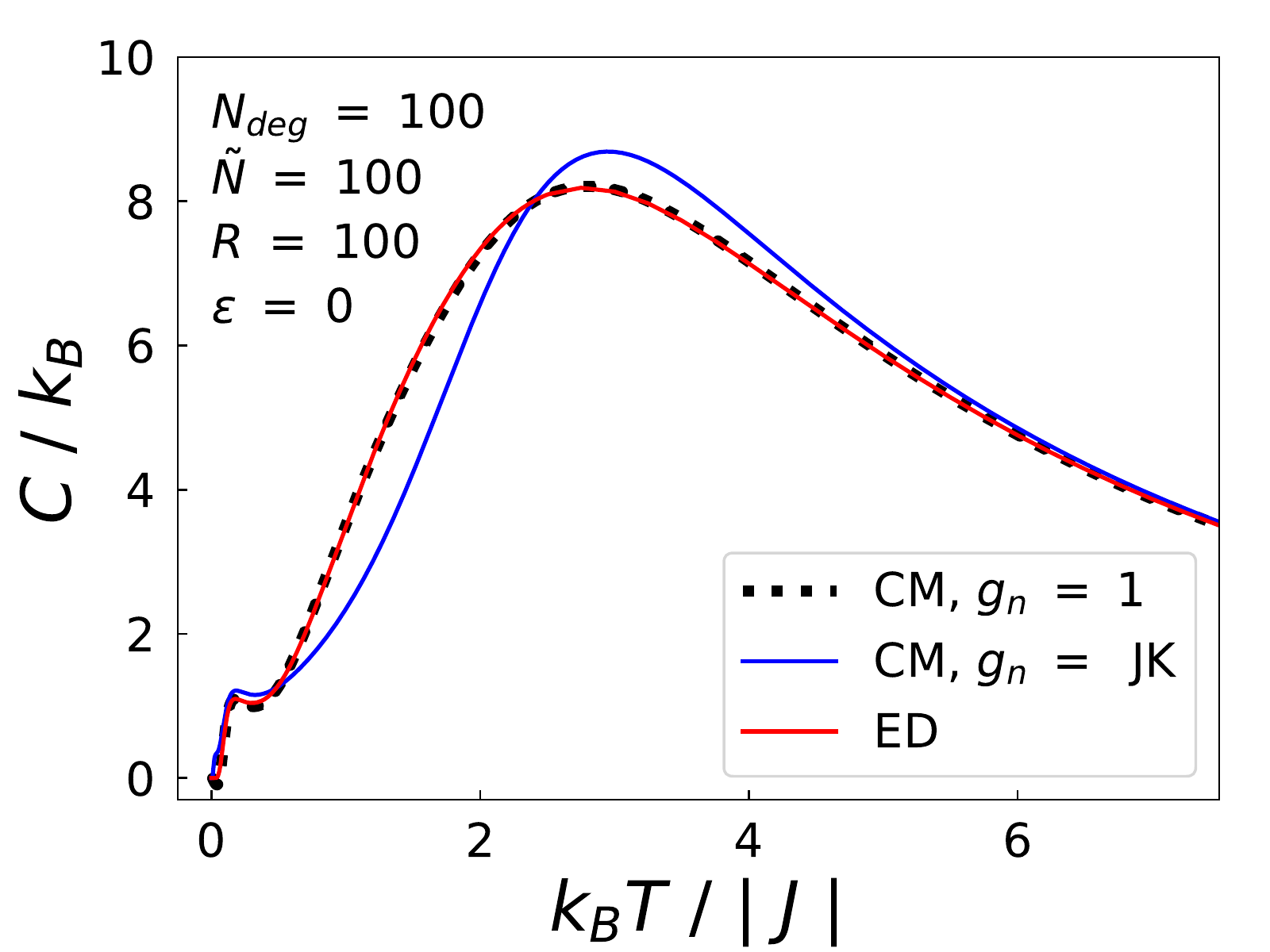}
 \includegraphics[width=0.69\columnwidth]{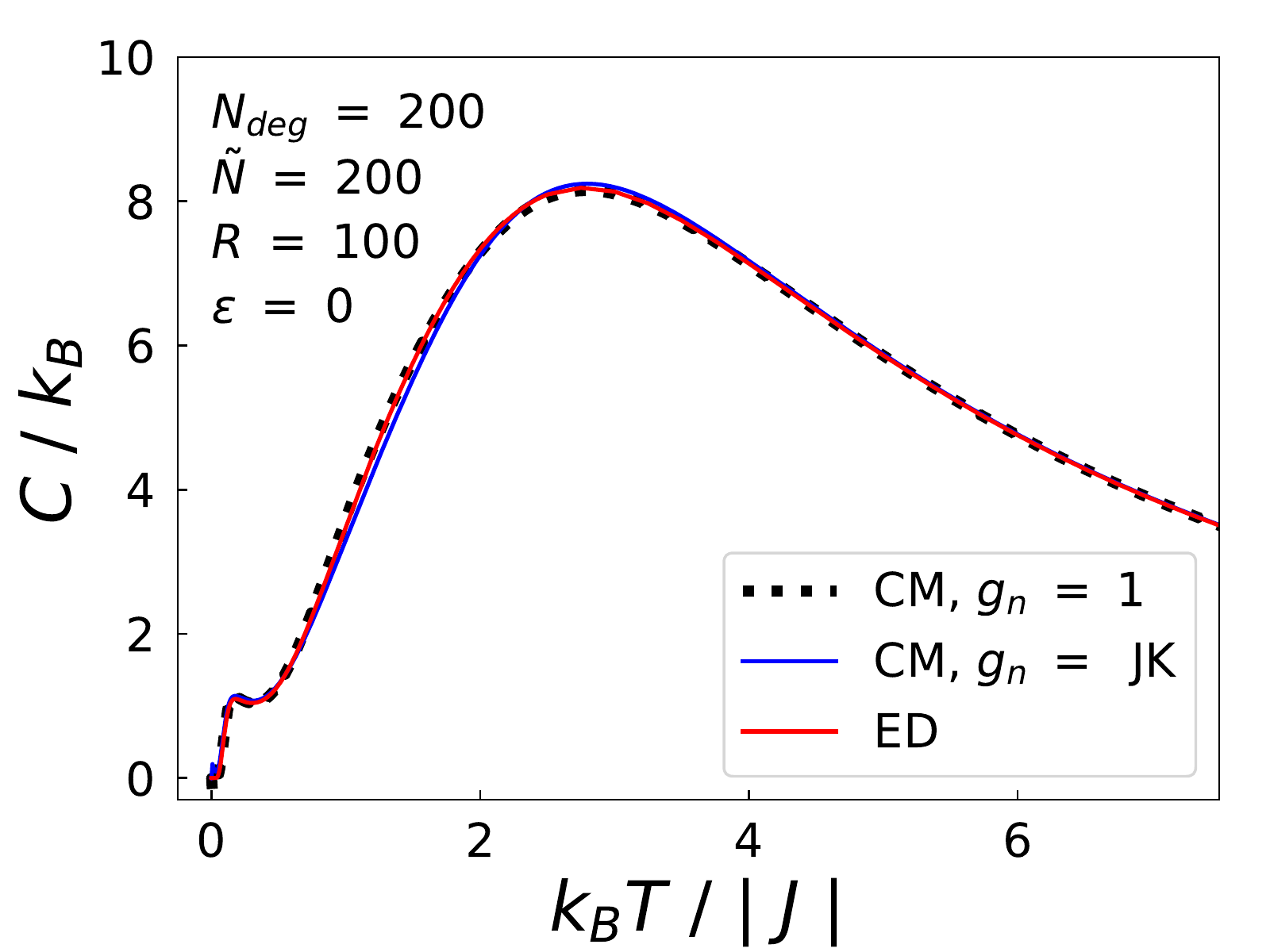}
 \caption{The heat capacity $C / k_B$ of the Heisenberg ring with $N=10$ spins $s=5/2$ computed using the Chebyshev method (CM) and exact diagonalization (ED). Displayed are the results with and without Jackson kernel (JK) for $\Ndeg = 100$ and $\Ndeg = 200$.}
 \label{C_j_ring}
 \end{figure}
\begin{figure}[ht!]
 \centering
 \includegraphics[width=0.69\columnwidth]{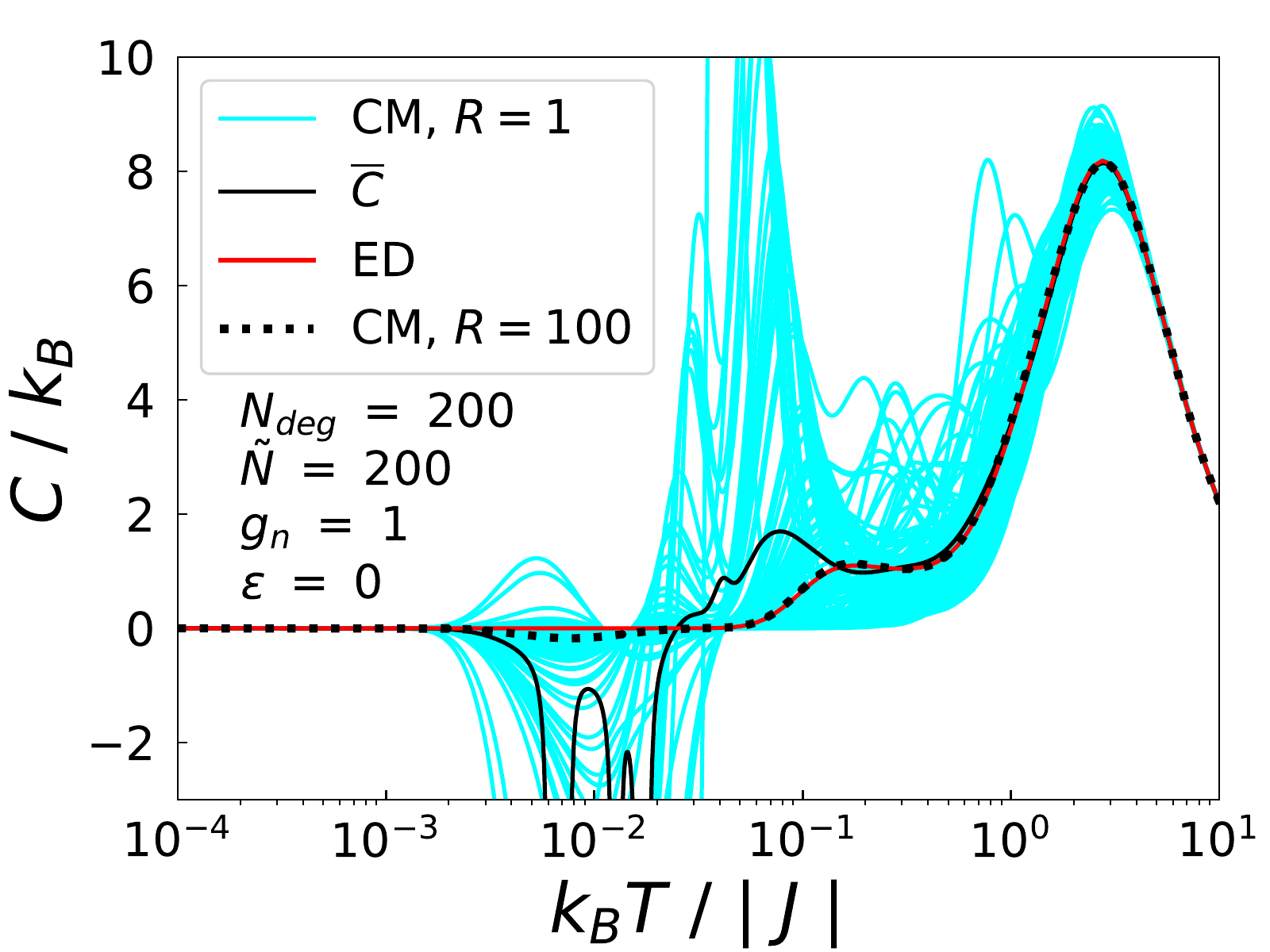}
  \includegraphics[width=0.69\columnwidth]{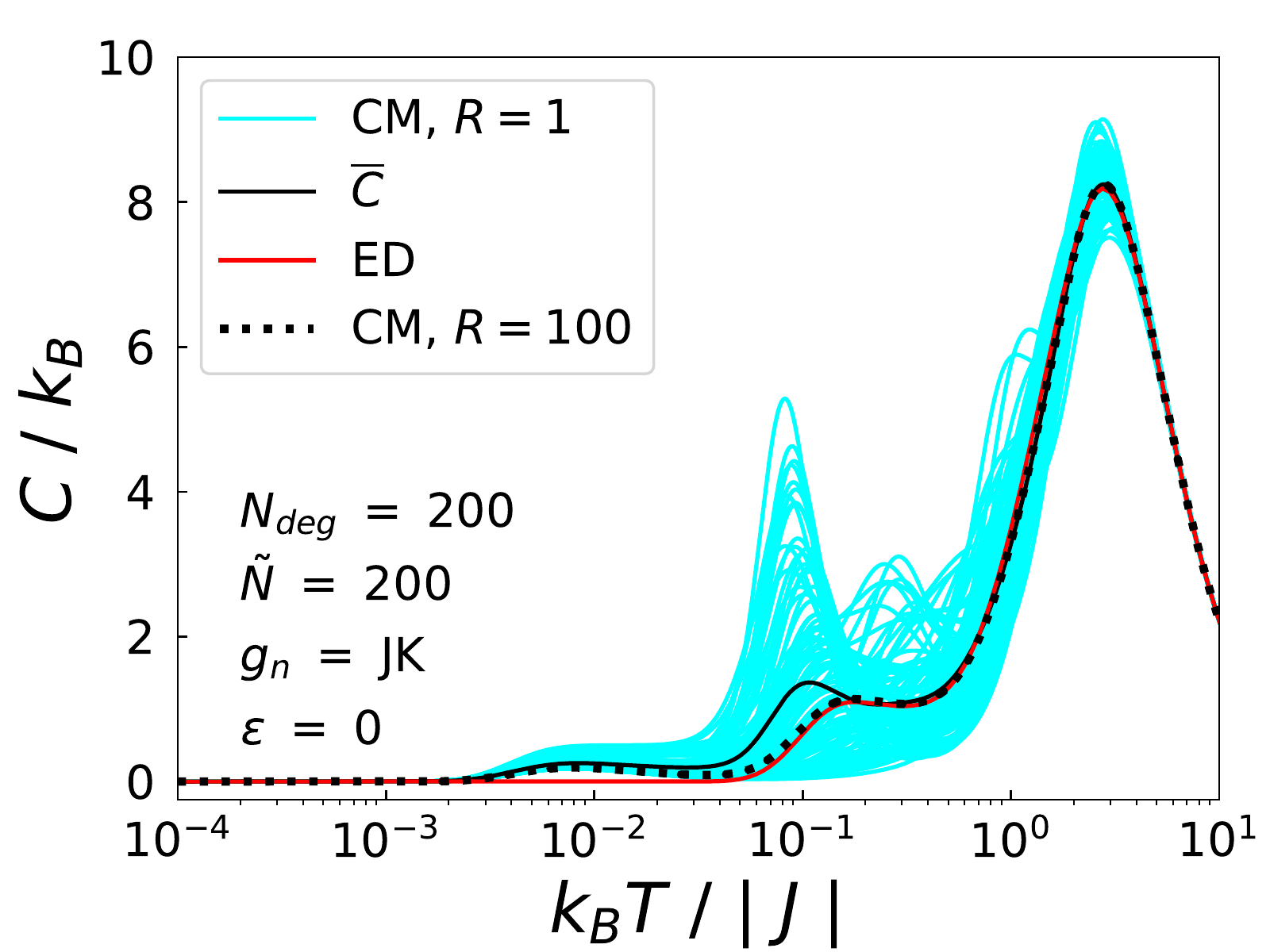}
 \caption{The heat capacity $C / k_B$ of the Heisenberg ring with $N=10$ spins $s=5/2$  computed using the Chebyshev method (CM) and exact diagonalization (ED). Displayed are the results of $P = 100$ realizations with $R=1$ and one with $R=100$; $\Ndeg=200$.
 Compare \figref{C_R_ring}.
 \label{C_R2_ring}}
 \end{figure}

The antiferromagnetic Heisenberg ring with $N=10$, 
$s=5/2$ and nearest neighbor interaction is an interesting 
example as well, as this system is realized as a magnetic molecule (abbreviated Fe$_{10}$) called the ``ferric wheel'' \cite{SRS:PRR20} that can be accurately described by this model. 
In \figref{C_R_ring} $P=100$ estimates with the Chebyshev method using $R=1$ random 
vectors and their mean are compared to an estimate using $R=100$ random vectors. 
They are displayed with and without kernel.

One can see that without kernel the estimates with $R=1$ are broadly scattered at low temperatures while their mean and the estimate with $R=100$ random vectors are almost perfectly aligned with the exact result. When employing the kernel the estimates with $R=1$ are distributed less broadly but their mean and the estimate with $R=100$ deviate strongly from the result of the exact diagonalization. From the experience collected before, we assume that these deviations can be resolved by using higher orders of expansion $\Ndeg$, see \figref{C_j_ring} with a linear temperature axis.
Note that even for $\Ndeg=200$ the result without kernel outperforms the one with kernel. Further more, the result with $\Ndeg=100$ without kernel is more accurate than the result with $\Ndeg=200$ and kernel.
   
However, because of the narrower distribution of the estimates when using the kernel we want to investigate the statistical behavior of the results obtained with a higher order of expansions, see \figref{C_R2_ring} for the results for $\Ndeg=200$. One can see that the narrowing of the $R=1$ estimates by using the kernel is less significant than in the $\Ndeg=100$ case but still non-negligible especially in the low temperature regime. This can be seen in the deviation of mean from the exact result which is greater in the case without kernel. But again the deviation is still there for the same and even slightly higher temperatures. The best result is obtained with the $R=100$ estimates without kernel. In this case an order of expansion of $\Ndeg=100$ is sufficient.

Also for the differential susceptibility (not shown) we obtain that the mean of the $R=1$ estimates 
deviates from the exact diagonalization result without kernel more strongly than with kernel. 
The $R=100$ estimate on the other hand is almost accurate without kernel 
but deviates as strongly as the mean in the case with kernel.

\begin{figure}[ht!]
 \centering
 \includegraphics[width=0.69\columnwidth]{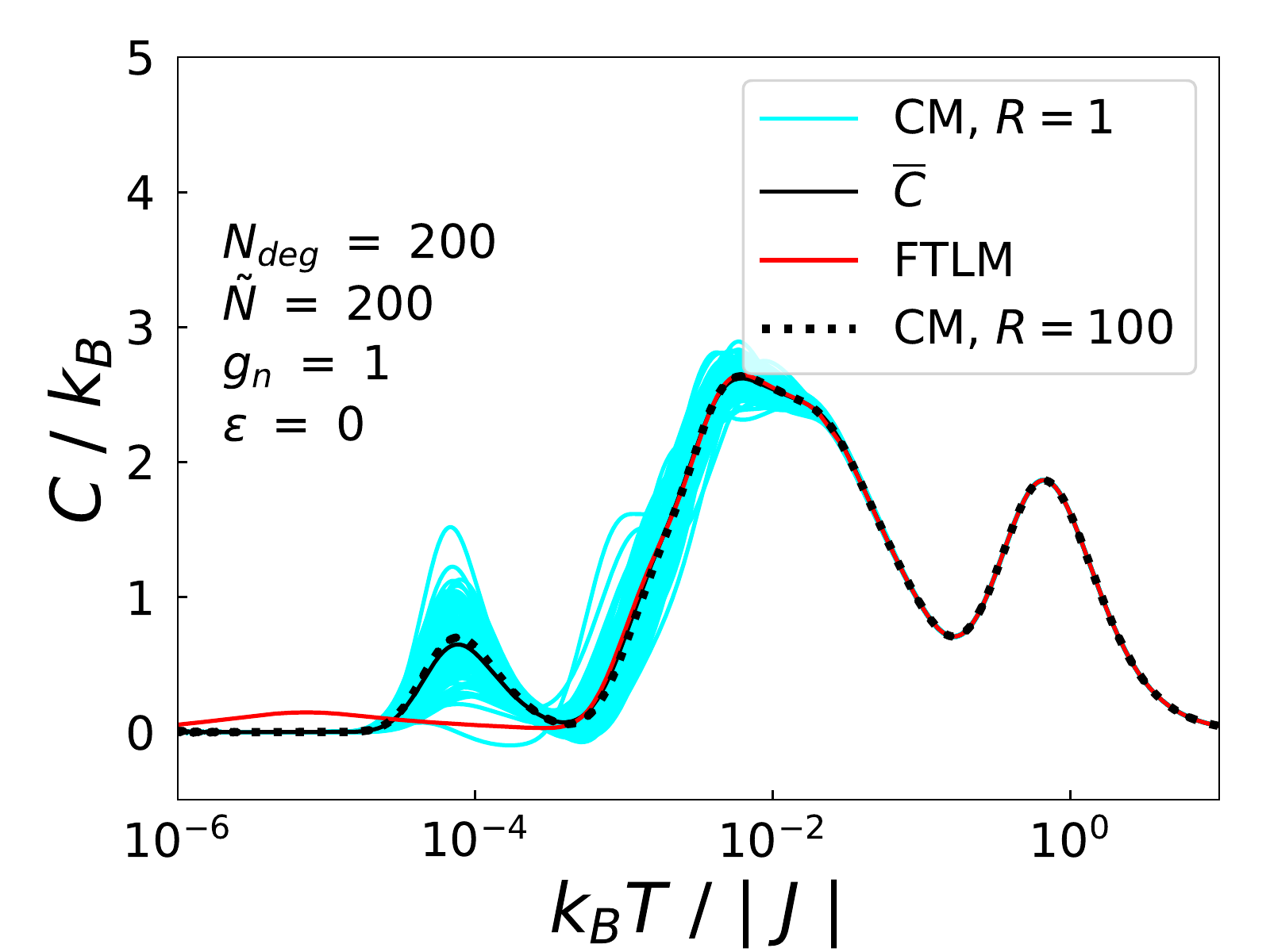}
  \includegraphics[width=0.69\columnwidth]{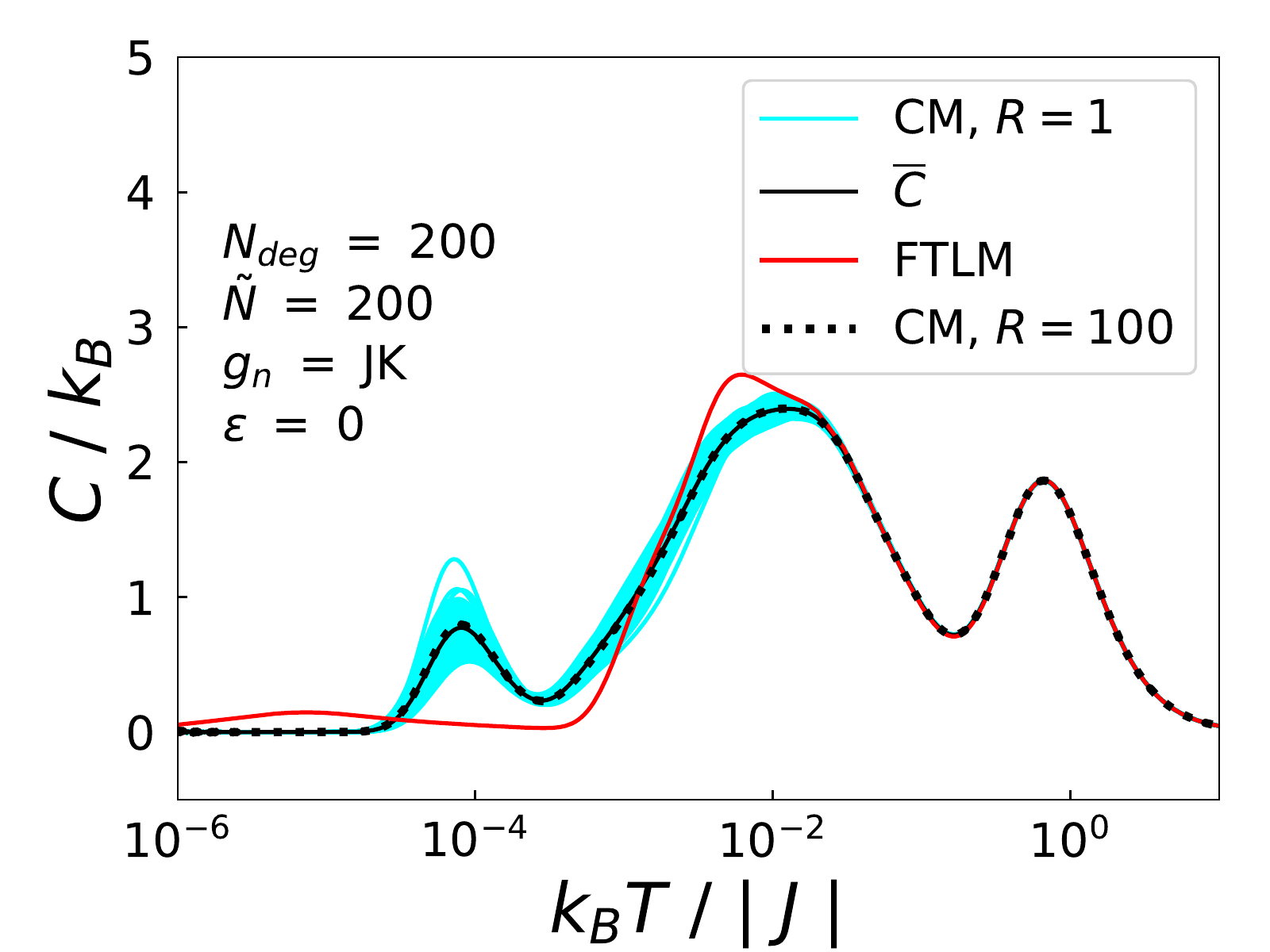}
 \caption{The heat capacity $C / k_B$ of the sawtooth chain with $N=24$ spins $s=1/2$  
 computed using the Chebyshev method (CM) the finite temperature Lanczos method (FTLM). 
 Displayed are the results of $P = 100$ realizations with $R=1$ and one with $R=100$.\label{C_R_chain}}
 \end{figure}

 \begin{figure}[ht!]
 \centering
 \includegraphics[width=0.69\columnwidth]{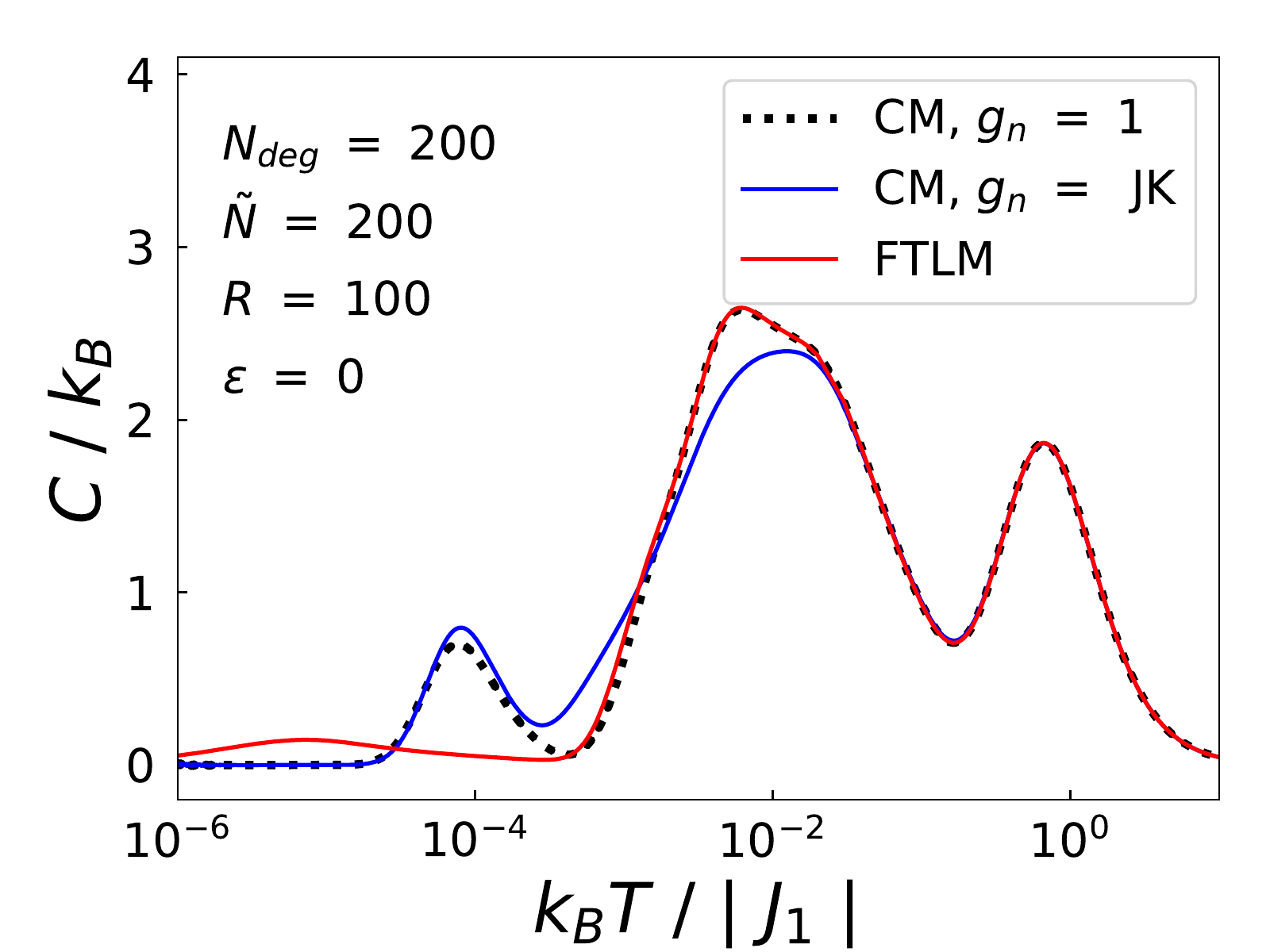}
 \includegraphics[width=0.69\columnwidth]{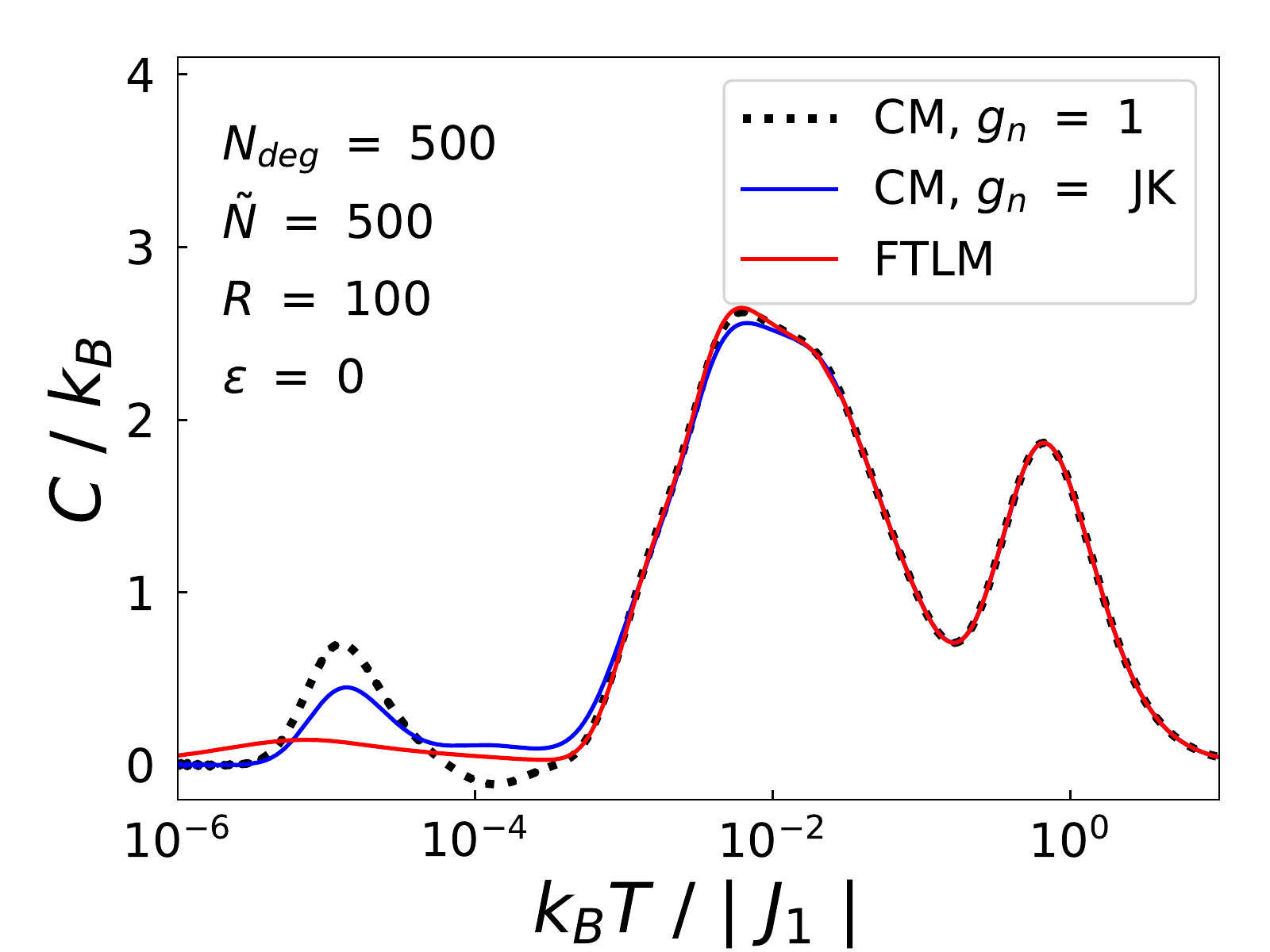}
 \caption{The heat capacity $C / k_B$ of the delta chain with $N=24$ spins $s=1/2$ computed using the Chebyshev method (CM) and the finite temperature Lanczos method (FTLM). Displayed are the results with and without Jackson kernel (JK) for $\Ndeg = 100$ and $\Ndeg = 500$.\label{C_j_chain}}
 \end{figure}

\subsection{Sawtooth chain}

The sawtooth chain (also known as delta chain) is an example with a highly degenerate spectrum.
The Hamiltonian reads
\begin{align}
\op{H} = J_1\sum_{i=1}^{N} \op{\vec{s}}_{i}\cdot \op{\vec{s}}_{i+1} 
+ J_2\sum_k^{N/2}\op{\vec{s}}_{2k-1}\cdot\op{\vec{s}}_{2k+1}     
\end{align}
with periodic boundary conditions, ferromagnetic nearest neighbor interaction $J_1<0$ 
and antiferromagnetic next-nearest neighbor interaction $J_2>0$.
We select a case with $|J_2/J_1|=0.45$ which is close to the quantum critical point (QCP)
at $|J_2/J_1|=1/2$ \cite{KDN:PRB14}.
The typicality approach has shown to be very efficient for this systems in schemes 
such as FTLM \cite{SRS:PRR20}. This  can be confirmed for the Chebyshev method as well.
In \figref{C_R_chain} the $R=1$ estimates of the heat capacity are distributed very narrowly around their mean which itself is perfectly aligned with $R=100$. While the result without kernel is also aligned with the FTLM estimate,
the result with kernel shows significant deviations from the FTLM result. 

We again show another result for a higher order of expansion $\Ndeg=500$, 
the lower graph in \figref{C_j_chain}. The deviation due to the kernel can be minimized, 
but still does not fall below those of the results without kernel.

\section{Discussion and conclusions}
\label{sec-4}
We have seen that the Chebyshev method achieves very accurate results when handled with care. 
There are many possible choices for the parameters introduced for this method.
We tried to identify some ``good" choices and some methods to optimize them. 

In particular we found that the number of points of integration $\tilde{N}$ 
should be chosen closely to the order of expansion which itself has to be 
chosen according to the dimension of the problem. In the cases investigated, 
$\Ndeg=100-200$ is a sufficient choice as well as $R\geq100$. 
The parameter $\varepsilon$ had no positive effect at least not when 
trying to approximate thermodynamic functions. 
So it seems advisable to set it equal to zero. 

The most interesting ``parameter'' was whether to smooth the result 
with the Jackson kernel or not. We found here as well that for the 
investigated systems there was no positive effect. 
However, there might be a different use of the kernel.
When the expansion of the density of states is completed the kernel can be employed 
without great computational effort to check the approximation with and without kernel
for differences. 
For good results with a high order of the expansion the kernel changes the results only 
where they were already wrong, e.g.\ where heat capacity is negative. 
Therefore, if the kernel does not change the result too much 
one can be reasonably sure that the choice of the order of expansion is sufficient.

Finally, we can summarize that the approach via Chebyshev polynomials is accurate,
but does not show any advantage compared to FTLM \cite{SRS:PRR20}.

\section*{Acknowledgment}

This work was supported by the Deutsche Forschungsgemeinschaft DFG
(355031190 (FOR~2692); 397300368 (SCHN~615/25-1); 449703145 (SCHN~615/28-1)). 
Computing time at the Leibniz Supercomputing Center in Garching is gratefully
acknowledged. 


%

\end{document}